\documentclass[twocolumn,twocolappendix,times]{aastex631}

\usepackage{comment}
\usepackage{amssymb}
\usepackage{aas_macros}
\usepackage{mathtools}
\usepackage[version=4]{mhchem}
\usepackage{multirow}
\usepackage{xspace}
\usepackage{threeparttable}
\usepackage{mathrsfs}
\usepackage{threeparttablex}

\newcommand{\pyird}{{\sf PyIRD}\xspace}
\newcommand{\ExoJAX}{{\sf ExoJAX}\xspace}

\accepted{\today}
\submitjournal{AJ}

\begin{document}

\title{Unveiling the Atmosphere of HR 7672 B from the Near-Infrared High-Resolution Spectrum Using REACH/Subaru}

\correspondingauthor{Yui Kasagi}
\email{kasagi.y96@gmail.com}

\author[0000-0002-8607-358X]{Yui Kasagi}
\affiliation{Institute of Space and Astronautical Science, Japan Aerospace Exploration Agency, \\3-1-1 Yoshinodai, Chuo-ku, Sagamihara, Kanagawa, 252-5210, Japan}
\affiliation{Department of Astronomical Science, School of Physical Sciences, The Graduate University for Advanced Studies, SOKENDAI, \\2-21-1 Osawa, Mitaka, Tokyo 181-8588, Japan}

\author[0000-0003-3800-7518]{Yui Kawashima}
\affiliation{Department of Astronomy, Graduate School of Science, Kyoto University, Kitashirakawa Oiwake-cho, Sakyo-ku, Kyoto 606-8502, Japan}

\author[0000-0003-3309-9134]{Hajime Kawahara}
\affiliation{Institute of Space and Astronautical Science, Japan Aerospace Exploration Agency, \\3-1-1 Yoshinodai, Chuo-ku, Sagamihara, Kanagawa, 252-5210, Japan}
\affiliation{Department of Astronomy, Graduate School of Science, The University of Tokyo, 7-3-1 Hongo, Bunkyo-ku, Tokyo 113-0033, Japan}

\author[0000-0001-6181-3142]{Takayuki Kotani}
\affiliation{Astrobiology Center, 2-21-1 Osawa, Mitaka, Tokyo 181-8588, Japan}
\affiliation{Astronomical Science Program, The Graduate University for Advanced Studies, SOKENDAI, 2-21-1 Osawa, Mitaka, Tokyo 181-8588, Japan}
\affiliation{National Astronomical Observatory of Japan, 2-21-1 Osawa, Mitaka, Tokyo 181-8588, Japan}

\author[0000-0003-1298-9699]{Kento Masuda}
\affiliation{Department of Earth and Space Science, Osaka University, Osaka 560-0043, Japan}

\author[0000-0002-1094-852X]{Kyohoon Ahn}
\affiliation{Subaru Telescope, National Astronomical Observatory of Japan, 650 North A‘ohōkū Place, Hilo, HI 96720, USA}
\affiliation{Technology Center for Astronomy and Space Science, Korea Astronomy and Space Science Institute, \\776 Daedeok-daero, Yuseong-gu, Daejeon 34055, Republic of Korea}

\author[0000-0002-1097-9908]{Olivier Guyon}
\affiliation{Subaru Telescope, National Astronomical Observatory of Japan, 650 North A‘ohōkū Place, Hilo, HI 96720, USA}
\affiliation{Astrobiology Center, 2-21-1 Osawa, Mitaka, Tokyo 181-8588, Japan}
\affiliation{Steward Observatory, The University of Arizona, Tucson, AZ 85721, USA}

\author[0000-0003-3618-7535]{Teruyuki Hirano}
\affiliation{Astrobiology Center, 2-21-1 Osawa, Mitaka, Tokyo 181-8588, Japan}
\affiliation{National Astronomical Observatory of Japan, 2-21-1 Osawa, Mitaka, Tokyo 181-8588, Japan}
\affiliation{Astronomical Science Program, The Graduate University for Advanced Studies, SOKENDAI, 2-21-1 Osawa, Mitaka, Tokyo 181-8588, Japan}

\author[0000-0001-5213-6207]{Nemanja Jovanovic}
\affiliation{Department of Astronomy, California Institute of Technology, Pasadena, CA 91125, USA}

\author[0000-0002-4677-9182]{Masayuki Kuzuhara}
\affiliation{Astrobiology Center, 2-21-1 Osawa, Mitaka, Tokyo 181-8588, Japan}
\affiliation{National Astronomical Observatory of Japan, 2-21-1 Osawa, Mitaka, Tokyo 181-8588, Japan}

\author{Julien Lozi}
\affiliation{Subaru Telescope, National Astronomical Observatory of Japan, 650 North A‘ohōkū Place, Hilo, HI 96720, USA}

\author[0000-0002-6510-0681]{Motohide Tamura}
\affiliation{Astrobiology Center, 2-21-1 Osawa, Mitaka, Tokyo 181-8588, Japan}
\affiliation{National Astronomical Observatory of Japan, 2-21-1 Osawa, Mitaka, Tokyo 181-8588, Japan}
\affiliation{Department of Astronomy, Graduate School of Science, The University of Tokyo, 7-3-1 Hongo, Bunkyo-ku, Tokyo 113-0033, Japan}

\author[0000-0002-6879-3030]{Taichi Uyama}
\affiliation{Department of Physics and Astronomy, California State University Northridge, 18111 Nordhoff Street, Northridge, CA 91330, USA}

\author[0000-0003-4018-2569]{Sebastien Vievard}
\affiliation{Subaru Telescope, National Astronomical Observatory of Japan, 650 North A‘ohōkū Place, Hilo, HI 96720, USA}
\affiliation{Astrobiology Center, 2-21-1 Osawa, Mitaka, Tokyo 181-8588, Japan}
\affiliation{Space Science and Engineering Initiative, College of Engineering, Institute for Astronomy, University of Hawaii, \\640 North Aohoku Place, Hilo, HI, 96720, USA}

\author[0000-0003-2754-9856]{Kenta Yoneta}
\affiliation{National Astronomical Observatory of Japan, 2-21-1 Osawa, Mitaka, Tokyo 181-8588, Japan}

\begin{abstract}

Characterizing the atmospheres of exoplanets and brown dwarfs is crucial for understanding their atmospheric physics and chemistry, searching for biosignatures, and investigating their formation histories. 
Recent advances in observational techniques, combining adaptive optics with high-resolution spectrographs, have enabled detailed spectroscopic analysis for directly imaged faint companions.
In this paper, we report an atmospheric retrieval on the L-type brown dwarf HR~7672~B using a near-infrared high-contrast high-resolution spectrograph, REACH (Y, J, H band, $R\sim100,000$), which combines SCExAO with IRD at the Subaru Telescope. 
Our model, developed based on the \ExoJAX spectrum code, simultaneously accounts for several factors, including the presence of clouds in the L dwarf's atmosphere as well as contamination from the host star's light and telluric absorption lines in the observed spectra.
Our analysis identified \ce{H2O} and \ce{FeH} as the primary absorbers in the observed J- and H-band spectra.
Additionally, the observed features were reproduced with a model that includes cloud opacity, assuming an optically thick cloud at the pressure $P_\mathrm{top}$.
The resulting temperature at the cloud top pressure suggests the potential formation of clouds composed of \ce{TiO2}, \ce{Al2O3}, or \ce{Fe}.
This study is the first science demonstration for faint spectra obtained by REACH, providing a foundation for future investigations into the atmospheres of exoplanets and brown dwarfs.

\end{abstract}


\section{Introduction}

The atmospheres of brown dwarfs are very complex, similar to those of Jupiter and gas giant exoplanets.
For example, thermochemical equilibrium favors the formation of molecules, and condensate clouds can even form in their cool atmospheres \citep[e.g., ][]{2015ARA&A..53..279M, 10.1146/annurev-astro-081817-051846}.
Characterizing brown dwarf atmospheres is a practical step toward advancing our understanding of exoplanet atmospheres, as the physical and chemical processes in their atmospheres share many similarities.
Additionally, brown dwarfs can be observed with a higher signal-to-noise ratio (S/N) than fainter planets.
In particular, ``benchmark'' brown dwarfs have played a crucial role in testing theoretical models of both their atmospheres and evolution.
These brown dwarfs orbit bright main-sequence stars, which allows key properties such as chemical composition and age to be inferred directly from their host stars, rather than relying solely on model estimates.

These cool atmospheres can be studied using various methods, such as transit spectroscopy, high-resolution Doppler spectroscopy, and direct-imaging spectroscopy \citep[cf. ][]{10.1146/annurev-astro-081817-051846}.
The direct-imaging spectroscopy has constrained atmospheric profiles and detected molecular species such as \ce{H2O}, \ce{CO}, \ce{CH4}, and \ce{NH3} with high confidence in the atmospheres of various substellar objects, including exoplanets such as HR 8799 system \citep[e.g.,][]{10.1126/science.1166585, 10.1088/0004-637X/733/1/65, 10.1126/science.1232003, 10.1088/0004-637X/804/1/61, 10.3847/1538-3881/aabcb8} and 51 Eri b \citep{10.1126/science.aac5891} due to the higher S/N compared to other methods.

However, these observations have been mainly performed using low-resolution spectrographs with wavelength resolutions ranging from several tens to a few thousand.
Such spectra generally reveal the atmosphere's deepest layer through continuum emission that can be modified by upper-layer opacity sources like clouds.
There are possible degeneracies in interpreting the temperature-pressure (T-P) profile and chemical abundances mainly due to uncertainties in cloud composition and behavior \citep[e.g.,][]{10.1016/j.geomorph.2016.12.029}.

Recently, the development of near-infrared high-resolution spectrographs and the establishment of observation techniques combined with Adaptive Optics (AO) have made it possible to obtain high-resolution spectra of such faint substellar companions.
These spectra have led to the clear identification and precise determination of the relative abundances of a variety of molecular species in the atmospheres of gas giants and brown dwarfs \citep[e.g.,][]{2021AJ....162..148W, 10.3847/1538-3881/ac56e2, 10.3847/1538-3881/ac9f19, 10.3847/1538-4357/ac8673, 2024ApJ...971....9H, 2025ApJ...988...53K}. 
It is also possible to investigate the T-P profile and a wide range of pressures above cloud tops from the line depth ratio or the pressure-broadening of absorption lines.

\subsection{REACH: Capabilities and Observational Challenges}

We observed one of the benchmark brown dwarfs, HR~7672~B, by using REACH \citep[Rigorous Exoplanetary Atmosphere Characterization with High dispersion coronagraphy,][]{2020SPIE11448E..78K} on the Subaru Telescope.
It is a combination of InfraRed Doppler \citep[IRD,][]{2012SPIE.8446E..1TT, 10.1117/12.2311836} and Subaru Coronagraphic Extreme AO \citep[SCExAO,][]{2015PASP..127..890J}, achieving the highest resolution ($R\sim100,000$) in the Y, J, and H bands. 
This work is the first attempt at atmospheric retrieval analysis for a REACH spectrum for a very high-contrast system with a faint companion.

The resolution and wide wavelength range make REACH a unique and complementary instrument among other high-resolution spectrographs fed by AO corrections, such as KPIC on Keck Telescope \citep[K band, $R \sim 35,000$; ][]{10.1117/12.2314037, 10.48550/arXiv.1909.04541, 10.1117/12.2562836} and HiRISE at the Very Large Telescope (VLT) \citep[H band, $R \sim 100,000$;][]{10.1051/0004-6361/202038517}.
This broad and complementary wavelength coverage is crucial for atmospheric retrieval analysis, as different wavelength ranges are capable of probing various altitudes within the atmospheric layer, offering a comprehensive understanding of atmospheric composition and dynamics.

Injecting light from off-axis objects is a major challenge for REACH observations, particularly for faint targets like exoplanets. 
Since fiber positions cannot be directly set based on the image from the SCExAO internal fast IR detector (C-RED2), accurate correction of image plane distortion is crucial. 
In \citet{2020SPIE11448E..78K}, binary systems with well-known orbits were used to calibrate focal plane distortion, but their contrast ratios were limited (H-band: $\sim$4, J-band: $\sim$10). 
Consequently, instrumental effects on faint target spectra, systematic errors, and speckle halo light contamination remained poorly understood. 
This study aims not only to characterize the atmosphere of a brown dwarf but also to address these issues for the first time.

\subsection{HR~7672~A and B}

HR~7672~A is a nearby G0V star \citep{10.3847/0067-0049/225/2/32}.
Its brown dwarf companion, HR~7672~B, was discovered at a separation of 0\farcs{8} using AO imaging from the Gemini and Keck Telescopes, with an estimated effective temperature of 1510 -- 1850 K and the spectral type of L4.5 \citep{10.1086/339845}.
Its apparent J-, H-, and Ks-band magnitudes are 
$14.39 \pm 0.20, \; 14.04 \pm 0.14, \; 13.04 \pm 0.10$, respectively \citep{10.1051/0004-6361:20031216}\footnote{The J-band flux is considered to be overestimated due to the huge speckle halo background in their measurement.}.
HR~7672~B orbits around the host star with high eccentricity and near edge-on orbit \citep[$e=0.50^{+0.01}_{-0.01}, \; i=97.3^{+0.4}_{-0.5}$; ][]{10.1088/0004-637X/751/2/97}.
Its dynamical mass of $M=72.7 \pm 0.8 \mathrm{M_{Jup}}$ was measured by combining relative astrometry, radial velocities, and accelerations from Gaia and Hipparcos proper motions \citep{10.3847/1538-3881/ab04a8}.
This mass places HR~7672~B to be near the boundary of stars and substellar objects \citep[e.g., ][]{2023A&A...671A.119C, 2000ApJ...542..464C, 1998A&A...337..403B}. 

From the K-band spectrum observed by NIRSPEC/Keck with a wavelength resolution of 1,400, \citet{10.1086/339845} identified strong absorption of \ce{H2O} and \ce{CO}, but no absorption from \ce{Na I}, which are features characteristic of L dwarfs \citep{2003ApJ...596..561M, 2005ApJ...623.1115C}. 
More recently, \citet{10.3847/1538-3881/ac56e2} obtained high-resolution ($R=35,000$) K-band spectra of HR~7672~B using NIRSPEC fed by KPIC.
Through a retrieval analysis with {\sf petitRADTRANS} \citep{10.1051/0004-6361/201935470}, they constrained the abundances of \ce{H2O} and \ce{CO} and provided upper limits on those of \ce{CH4} and \ce{CO2}. 
Additionally, they demonstrated that the composition of HR~7672~B is 1.5 $\sigma$ consistent with that of its host star. 
To date, these are the only published constraints on the atmospheric composition of this brown dwarf.

This paper is organized as follows.
In Section \ref{sec:exojax_observation}, we introduce REACH observations for HR~7672~B and the properties of the systematics in a REACH spectrum we found during our analysis.  
In Section \ref{sec:exojax_model}, we explain our retrieval model including the settings for the opacity calculation and additional models for the contaminating light from the host star, telluric transmittance, and clouds in the brown dwarf's atmosphere.
In Section \ref{sec:exojax_results}, we show the results of the retrieval, including the joint analysis of the J- and H-band spectra.
Section \ref{sec:exojax_discussion} discusses these results.
In Section \ref{sec:exojax_conclusion}, we summarize our retrieval results and future prospects.

\section{REACH Data Reductions}
\label{sec:exojax_observation}
\subsection{Observations}

\begin{figure}[]
  \begin{center}
     \includegraphics[width=0.8\linewidth]{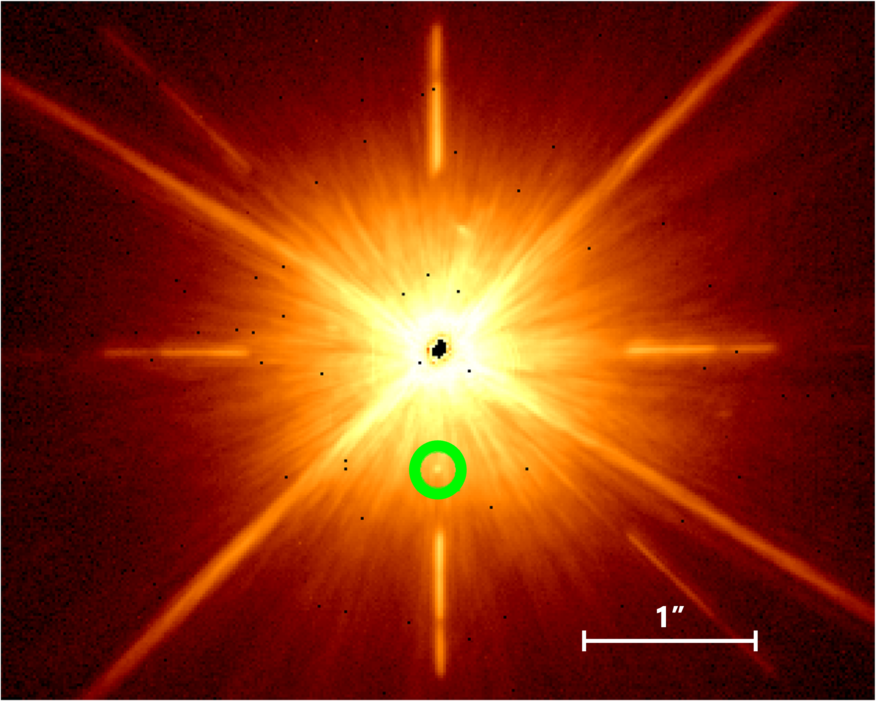}
  \end{center}
  \caption{Image of HR 7672 observed by SCExAO internal fast IR detector (C-RED2) on June 7, 2021 (ID: S21A-062, PI: H. Kawahara). HR~7672~B in a green circle is 0\farcs{7} away from the host star.}
  \label{fig:hr7672_image}
\end{figure} 

\begin{figure*}[]
  \gridline{\fig{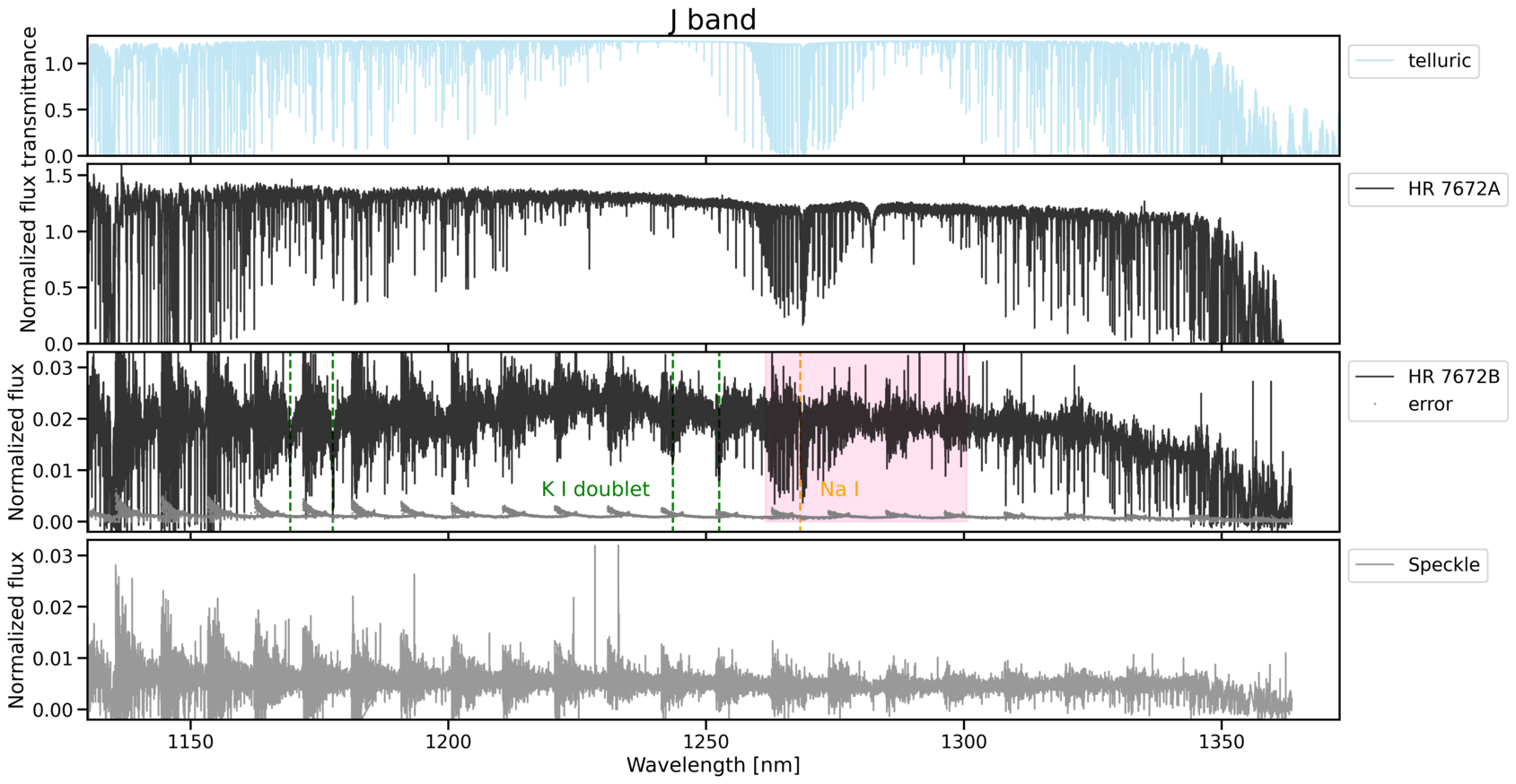}{\textwidth}{}}
  \gridline{\fig{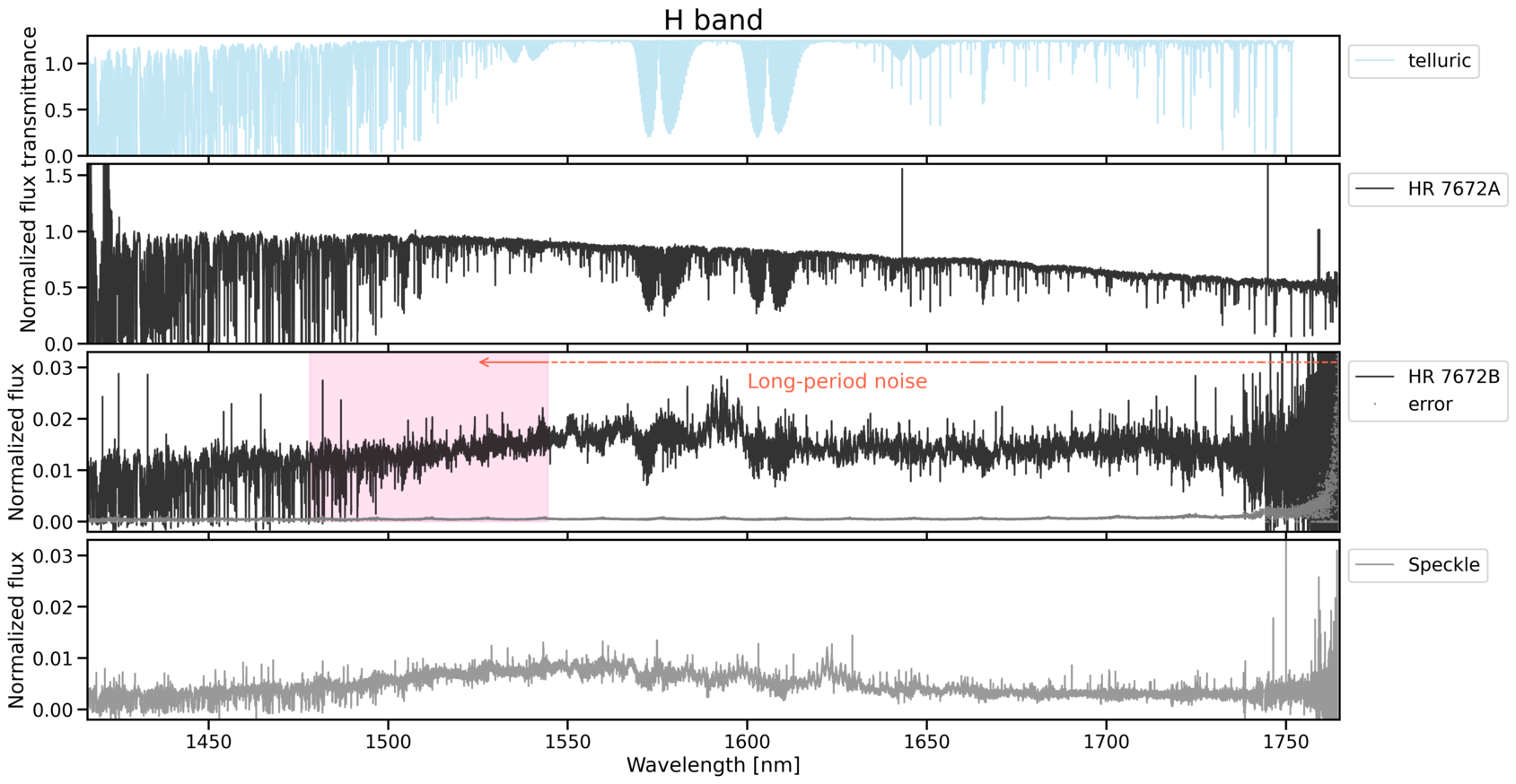}{\textwidth}{}}
  \caption{The observed J- and H-band spectra of HR~7672~B. The top panels of each figure show the telluric transmittance created by the LBLRTM \citep{10.1016/j.jqsrt.2004.05.058} for reference. The bottom three panels show the spectra of HR~7672~A, HR~7672~B, and the speckle halo light, with each combining two frames. The last spectrum was obtained from the speckle halo fiber during the observation of HR~7672~B. The gray dots in the third panel are uncertainties at each data point. Wavelength ranges used for our retrieval are highlighted in pink. The mean uncertainties within these ranges are $7.8 \times 10^{-4}$ for the J band and $8.1 \times 10^{-4}$ for the H band, respectively. These uncertainties are expressed as relative values with respect to the flux normalized by the blaze function. In the J band, green dashed lines are the wavelengths of \ce{K I} doublets, and the orange dashed line is at the absorption of \ce{Na I}. In the H band, wavelength regions in the red dashed line exhibit pronounced long-period noise. 
  }
  \label{fig:obs_hr7672b}
\end{figure*}

HR~7672~A and B were observed with REACH on June 6 and 24, 2021, respectively (ID: S21A-062, PI: H. Kawahara), with exposure times of 60 seconds for HR~7672~A and 1500 seconds for HR~7672~B. 
Ideally, the host star and the companion would be observed on the same night, but in this case, they were observed on separate nights due to scheduling and visibility constraints, partly resulting from the need to avoid persistence from the bright host star.
As discussed in Section~\ref{sec:leak_evaluate}, the observing conditions on both nights showed no significant differences, and the telluric lines in the spectra were also largely consistent.
An astrometry prediction was not required for fiber placement on this target, as its position was visually confirmed using the image obtained with SCExAO’s internal fast IR camera (C-RED2), as shown in Figure \ref{fig:hr7672_image}. 
Based on this image, the off-axis parabolic mirror (OAP) was adjusted to inject the target’s light into the fiber.

REACH has a bundle of 7 single-mode fibers (SMFs) arranged in a hexagonal shape with a fiber-to-fiber separation of 37 $\mu$m (159 mas).
Each SMF has a Mode Field Diameter (MFD) of 10.4 $\mu$m (44.8 mas), defined at a wavelength of 1550 nm.
The small fiber diameter enables observations of HR~7672~B, located approximately 0\farcs{7} from its host star.
The central fiber is designed to capture light from the scientific target, while the surrounding fibers, referred to as ``speckle halo fibers,'' collect residual scattered stellar light or light from the Laser Frequency Comb (LFC).
The central fiber and one of the speckle halo fibers are connected to IRD, with their respective outputs appearing on the Y/J- and H-band detectors.

Data reduction was performed using \pyird version 1.0.0 (Y. Kasagi et al., submitted)\footnote{\url{https://github.com/prvjapan/pyird}}, a newly developed open-source data reduction tool for IRD and REACH data. 
This pipeline performs standard procedures for echelle data reduction, including detector noise removal, flat-fielding, one-dimensional spectral extraction, and wavelength calibration using a Th-Ar lamp.
It is specifically designed to effectively remove noise on a detector and extract high-S/N spectra from faint targets.
The overall detector noise, including readout noise and other instrumental or background components, is modeled and subtracted using regions of the detector that are not illuminated by the science light.
The S/N ratios around 1.5 $\mu$m were 284 for HR~7672~A and 27 for HR~7672~B, both calculated from the combined data of the two observed frames. 
When the contribution from the host star's leaking light is taken into account, the S/N of the companion-only spectrum is reduced to 17. 

Figure \ref{fig:obs_hr7672b} shows the J- and H-band spectra for HR~7672~A and B, along with the spectra of a speckle halo light obtained simultaneously with the HR~7672~B observation. 
These spectra are normalized using blaze functions for each band. 
The Y-band spectrum is not shown due to its low S/N. 
Outliers, including hot pixels and bad pixels, were removed using a 5-sigma clipping method applied to each spectral order.
The pink highlighted regions in Figure \ref{fig:obs_hr7672b} indicate the wavelength ranges used in the retrieval analysis (see Section \ref{sec:exojax_model}). 

The wavelength range from 1570 to 1670 nm in the H band is affected by long-period noise, making it difficult to distinguish between noise and the scientific signal. 
As a result, this range was excluded from our analysis. 
However, understanding the nature of this noise is essential for future REACH observations and the development of next-generation spectrographs.
A detailed investigation of this noise is presented in Appendix \ref{sec:fringe_appendix}, while a summary of its properties is provided in Section \ref{sec:fringe}. 

\subsection{Systematics in a REACH Spectrum}
\label{sec:fringe}
\begin{figure*}[]
    \begin{center}
       \includegraphics[width=\linewidth]{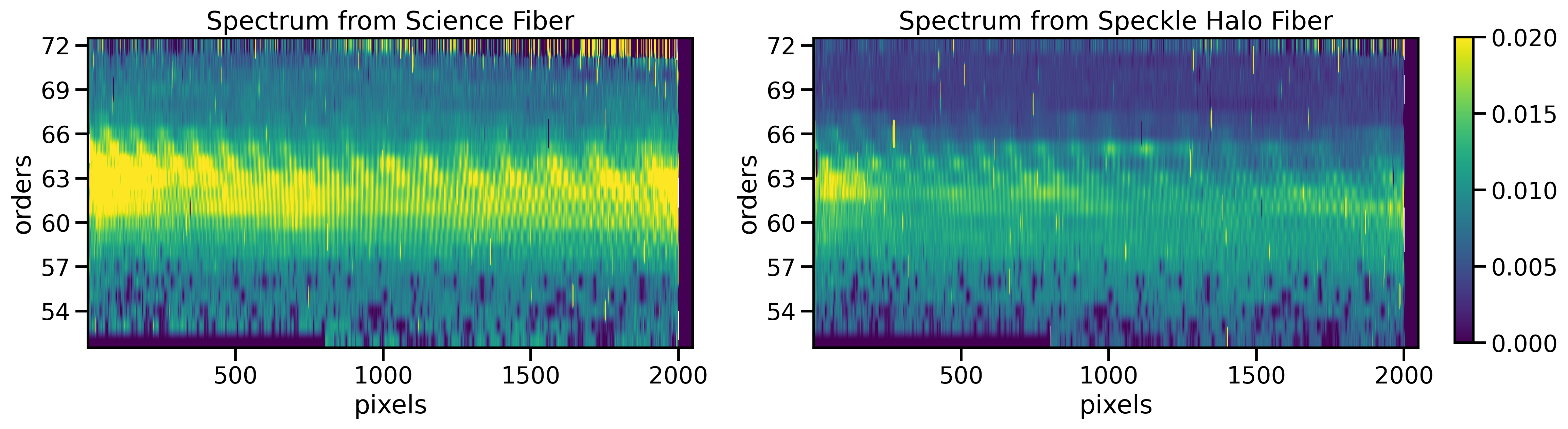}
    \end{center}
    \caption{Heatmap of the H-band spectrum of $\kappa$ And b, shown across spectral orders and pixels within each order. The color gradually changes in accordance with the normalized flux. The left panel is the spectrum obtained from the science fiber, while the right panel is the spectrum obtained from the speckle halo fiber. The periodic noise appears prominently especially from orders 61 to 66, and they look similar in both fibers. \label{fig:heatmap_h}}
\end{figure*}

Our analysis revealed two dominant types of noise: short-period and long-period variations. 
The short-period noise has a typical wavelength scale of approximately 0.18 nm, with amplitudes ranging from 0.67 \% to 1.3 \%, depending on the order\footnote{The term ``order'' in this paper does not mean the number of optical diffraction orders according to the characteristics of the diffraction grating. Instead, the numbering of ``order'' commences from 1 at the shortest wavelength found on the Y/J band detector, encompassing 51 orders (order 1 -- 51), and concludes at 72 on the H band detector's longest wavelength, spanning 21 orders (order 52 -- 72).}.
The long-period noise has a typical scale of approximately 1.2 nm and can reach amplitudes of around 15 \% to 22 \% at its peak.
This long-period noise is particularly evident in the H band, as illustrated in Figure \ref{fig:heatmap_h}.
The figure displays the spectrum of the brown dwarf $\kappa$ Andromedae b ($\kappa$ And b), which is most affected by the noise among our observed targets.
The color in the figure represents variations in the normalized flux. 
It reveals a common noise pattern in the spectra of the science and speckle halo fiber.

\begin{figure}[]
  \begin{center}
     \includegraphics[width=\linewidth]{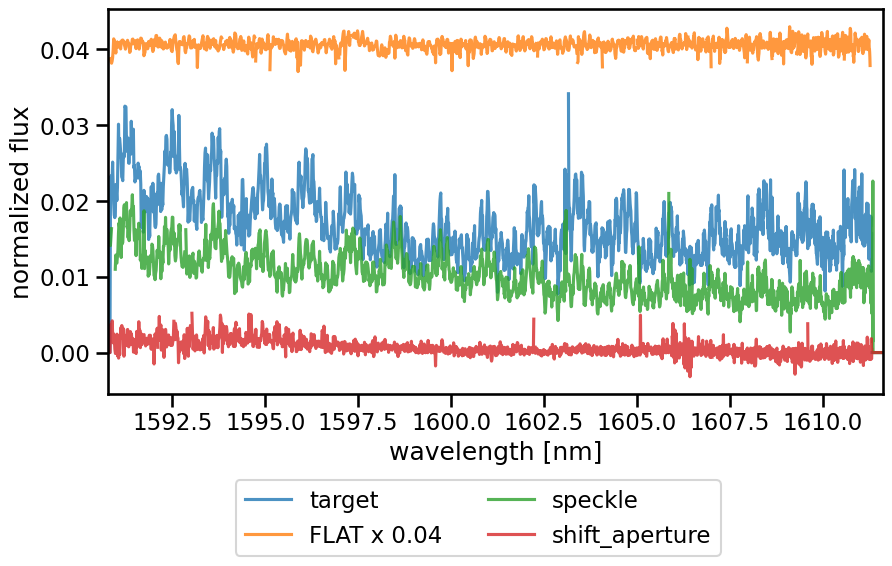}
  \end{center}
  \caption{The example of spectra of order 64 (in the H band) observed on the same date. The blue line is the spectrum of $\kappa$ And b, the orange line is the FLAT spectrum scaled by a factor of 0.04 for comparison, the green line is the spectrum from speckle halo fiber of the $\kappa$ And b observation, and the red line is the spectrum of $\kappa$ And b with shifting trace line by 40 pixels. See the text for details. \label{fig:fringe_spectra}}
\end{figure}

A long-period noise trend is evident over a wide wavelength range from order 63 to order 67 (1.57–1.67 $\mu$m), with the most pronounced trend occurring in order 64.
Figure \ref{fig:fringe_spectra} shows the normalized flux of the target spectrum in order 64. 
It also shows a spectrum of light from a white lamp used for flat fielding (hereafter FLAT spectrum), a spectrum from the speckle halo fiber, and a spectrum with no light.
The last spectrum was extracted by setting apertures shifted 40 pixels from the target light position in the spatial direction on the detector. 
This was used to examine whether the periodic noise appears across the detector.
Since no periodicity was found in the last spectrum, these periodicities are not additive but rather multiplicative to the spectrum from the fibers.

\subsubsection{Summary of Properties of Periodic Noise and Estimated Origin}
\label{sec:pyird_fringe_summary}
The following key properties of the noise were identified from our analysis:
\begin{itemize}
  \item Periodic noises are particularly prevalent in the spectra of faint targets observed with REACH. In particular, the long-period noise was not observed in the spectra of brighter targets (such as HR~7672~A, shown in Figure \ref{fig:HR7672A_HPaMg}), even when observed on the same night.
  \item There is a positive correlation between the periods of noises and the wavelength (order) for both shorter and longer periodicities, particularly in orders where the longer periodicity is apparently present.
  \item The short-period noise is evident in the FLAT spectrum as well as in the spectra from both the science and speckle halo fibers.
  \item The long-period noise does NOT manifest in the FLAT spectrum. However, this specific periodicity can be detected in the spectra of a faint target from both science and speckle halo fibers.
  \item While the periodicities generally remain stable regardless of the observation date, the amplitude of the longer periodicity might exhibit variations.
\end{itemize}

Although the exact cause of periodic noise is not fully understood, we estimate that it arises from instrumental effects, probably the interference related to the single-mode fibers (SMFs) that REACH used.
This conclusion is supported by the fact that short-period noise is absent in FLAT spectra taken by IRD, which uses multi-mode fibers.
Similar fringing signals have been observed in spectra from the KPIC/Keck instrument, where \citet{10.1117/12.3018020} has identified the noise sources as transmissive optics, including two dichroic mirrors and the spectrograph's entrance window.
However, the optical design of REACH, which consists of IRD and SCExAO, does not include a dichroic mirror, making this explanation unlikely.
Additionally, the interference between the speckle halo fiber and the science fiber can be excluded. 
The spectra of HR~7672~B still exhibit similar periodic noise, even when only target light was injected into the science fiber and no light into the speckle halo fiber.

The possibility remains that the noise stems from reflections at the fiber end face or the close proximity of SMFs in the REACH instrument. 
There are two possible solutions to remove the periodic noise: instrument modifications and post-processing of spectra. 
The first approach includes using Angled Physical Contact (APC) type SMFs to direct reflected light away from the fiber, or using a single SMF to avoid light leakage between fibers. 
The second approach involves modeling the noise with multiperiod sinusoidal functions or fitting it simultaneously with atmospheric retrieval using Gaussian Processes. 
Further examination is essential to pinpoint the exact cause of the noise, which will be crucial for future developments.

\section{Spectral Modeling}
\label{sec:exojax_model}
Our framework for modeling brown dwarf atmospheres is based on \ExoJAX version 1.5.1
\footnote{\url{https://github.com/HajimeKawahara/exojax}} 
\citep{10.3847/1538-4365/ac3b4d, 2025ApJ...985..263K}, which is the code to compute the autodifferentiable spectrum model for the high-dispersion characterization. 

In our analysis, we only used a part of the observed spectrum.
We did not use the Y-band spectrum due to a low S/N.
Additionally, wavelength ranges at the shorter end of the J-band and the longer end of the H-band were excluded to avoid the \ce{K I} doublet around $1.25\, \mu \mathrm{m}$ and the prominence of long-period noise, respectively. 
We will be able to use the entire J- and H-band spectra once we complete the verification of atomic line models in \ExoJAX and implement post-processing methods to mitigate long-period noise.

The fitting of the high-resolution spectrum was performed in relative flux units, with both the observed and model spectra normalized by the median flux around the central wavenumber ($\nu_0$) of a designated reference order.
Simultaneously, we computed the J-band photometric magnitude and fitted it to the magnitudes observed by \citet{10.1051/0004-6361:20031216}.

In the following subsections, we outline the foundational settings for modeling the spectrum of the brown dwarf (Section \ref{sec:model_opa}). 
Additionally, we discuss models uniquely added to fit the spectrum of HR~7672~B observed by REACH; these include considerations for contamination by telluric absorption lines (Section \ref{sec:model_telluric}), the presence of clouds in the brown dwarf's atmosphere (Section \ref{sec:model_cloud}), and contamination from light leakage from the host star (Section \ref{sec:leak}). 
Further parameters are introduced in Section \ref{sec:model_others}, and all free parameters are summarized in Table \ref{table:parameters_priors}.
The likelihood and HMC settings are outlined in Section \ref{sec:likelihood}.

\subsection{Opacity Calculations}
\label{sec:model_opa}
For the opacities of gaseous molecules, we consider line absorption by \ce{H2O} and \ce{FeH} for the J band, and only \ce{H2O} for the H band.
The volume mixing ratios of these molecules are assumed to be constant across all pressure levels considered.
We used the line lists of POKAZATEL \citep{2018MNRAS.480.2597P} and MoLLIST \citep{2003ApJ...594..651D, 2020JQSRT.24006687B} from Exomol \citep{10.1016/j.jms.2016.05.002} for \ce{H2O} and \ce{FeH}, respectively.
The contributions from other molecules, such as \ce{CH4}, \ce{CO}, and \ce{CO2}, are expected to be negligible in the wavelength range relevant to our retrievals, based on both the absorption cross-sections shown in \citet{2018arXiv180408149F} and the chi-square-based optimization in our analysis. 
For selecting the appropriate database for each molecule, we basically referred to \citet{10.1093/mnras/stac1412}, which validated the absorption lines of a T6 dwarf across both the H and K bands.
Additionally, we took into account the collision-induced absorption (CIA) by \ce{H2-H2} and \ce{H2-He} from HITRAN \citep{2019Icar..328..160K} as the continuum.

In our opacity computations, we determined the number of wavenumber bins ($N_\nu$) to correspond with the wavenumber resolution of the model, $R = 1,000,000$, which is set to ten times the resolution of REACH ($R_\mathrm{inst}=100,000$).
The model is convolved using both an Instrumental Profile (IP) broadening kernel and a rotation kernel.
The IP kernel was computed under the simple assumption of a Gaussian broadening, as expressed by
\begin{equation}
  k_{G}(\nu) = \frac{1}{\sqrt{2\pi \beta^2}}\exp{\left(-\frac{\nu^2}{2\beta^2}\right)},
\end{equation}
where $\beta$, the standard deviation, is defined as $\beta_{\mathrm{IP}}=c/(2\sqrt{2\ln{2}}R_\mathrm{inst})$.
During the fitting process, the model is resampled onto the same wavelength grid as the observed spectrum.
We adopted a simple temperature-pressure (T-P) profile by the power law index of $\alpha$, $T(P)=T_0 (P/P_0)^\alpha$, with 300 atmospheric layers, where $T_0$ is the temperature at a reference pressure $P_0$.
The reference pressure $P_0$ is set to $1\, \mathrm{bar}$ for models without cloud opacity, whereas, for models with cloud opacity, $P_0$ is defined as the pressure at the cloud top, $P_\mathrm{top}$.
In the case of cloudy models, this approach facilitates a clearer interpretation of the results, as the pressure contributing significantly to the opacity in the spectra is influenced by both the T-P profile and the cloud-top pressure.

For the computation of the J-band magnitude, we accounted for the filter transmission used in the Wide Field Camera for the UK Infrared Telescope on Mauna Kea \citep[WFCAM/UKIRT; ][]{10.1051/0004-6361:20066514} for our photometric computation.
The opacity calculation is performed with $R_\mathrm{inst}\sim7,000$, which is approximately the resolutions of the WFCAM filters, and $R=10\times R_\mathrm{inst}$.
Other parameters for opacity calculation are the same as the high-resolution model.

We calculated the opacities for each molecule by using {\sf PreMODIT} \citep{2025ApJ...985..263K}, a feature of \ExoJAX that pre-computes the lineshape density before applying temperature corrections and computes opacities using MODIT, a modified version of the discrete integral transform.

\subsection{Telluric Transmission Spectrum}
\label{sec:model_telluric}
The contamination of absorption lines from the Earth's atmosphere, so-called telluric lines, is one of the main noise sources, especially in the near-infrared \citep[e.g.,][]{10.1088/1538-3873/aac1b4, 10.1051/0004-6361/201833282}.
Completely removing telluric lines is challenging due to their variability depending on observational conditions, such as the water vapor content in the Earth's atmosphere and wind patterns.
One potential solution is to employ a wavelength mask, whereby masked regions are ignored.
However, this approach might result in a loss of approximately 15 \% of spectra in the Y, J, and H bands when rejecting telluric lines that have absorption levels of $> 5\,\%$  \citep{10.1117/12.2056385}.

Methods for correcting these lines can be roughly classified into empirical, data-driven, and forward-modeling approaches.
An empirical technique is the division by a telluric standard star's spectrum.
For this purpose, rapidly rotating B- to A-type stars are often used, as their absorption lines are broadened to the extent that one can assume only the telluric lines remain in their spectrum.
With a data-driven technique, stellar and telluric lines can be separated by applying principal component analysis (PCA) for many spectra taken on varying dates \citep[e.g.,][]{10.3847/1538-3881/ab40a7}.
A forward-modeling technique is modeling absorption lines through the computation of radiative transfer \citep[e.g.,][]{10.1051/0004-6361/201423932, 10.1051/0004-6361/201423909, 10.1088/0004-6256/148/3/53, 10.1051/0004-6361/201322383}.
This method does not require extra observational time for telluric standard stars or a comprehensive dataset for a specific target.

Because the above methods are necessary to distinguish between stellar and telluric lines prior to correction, we have chosen an alternative approach: simultaneous modeling of telluric transmittance, $T(\nu)$, using cross sections, $\sigma(\nu)$, as derived by \ExoJAX;
\begin{equation}
  \label{eq:transmittance}
  \mathrm{T}(\nu) = e^{-\sum_i \beta_i\sigma_i(\nu)},
\end{equation}
where $\beta_i$ represents the scaling coefficient for an absorber, and the subscript $i$ denotes specific molecular species.
We included the following primary molecular species as absorbers: \ce{H2O}, \ce{CO2}, and \ce{CH4} in the H band, and \ce{O2} in the J band \citep[cf.][]{1981aalt.book.....P, 10.1051/0004-6361/201423932}. 
The line lists used are as follows: POKAZATEL \citep{2018MNRAS.480.2597P} from ExoMol for \ce{H2O}, UCL-4000 \citep{2020MNRAS.496.5282Y} from ExoMol for \ce{CO2}, \ce{CH4} from HITEMP \citep{2020ApJS..247...55H}, and \ce{O2} from HITRAN \citep{2022JQSRT.27707949G}.
Their opacity is calculated using the {\sf Direct LPF} method, which directly calculates the line profile function (LPF) in \ExoJAX.
The pressure and temperature are fixed to the values at the summit of Mauna Kea, 0.6005 bar and 273 K, respectively.
Furthermore, we included the Doppler shift of the telluric absorptions ($v_\mathrm{tel}$) as a parameter, as the positions of the observed lines may shift slightly due to atmospheric winds.
Typical wind speeds at the summit of Mauna Kea are approximately 5–7 $\mathrm{m\,s^{-1}}$ \citep{2022PASP..134i5001V}.

\subsubsection{Validations}

We verified our modeling approach described by Equation \eqref{eq:transmittance} using the spectrum of the telluric standard star HR 7596 \citep[B9IV: ][]{1999MSS...C05....0H}, observed with REACH on June 6, 2021.
The fitted model shows minor discrepancies in the line profiles compared to the observed spectrum, with differences of up to 0.1 in the normalized spectrum around several telluric lines. 
These discrepancies are considered to have no significant impact on the retrieval analysis, as they are comparable to the spectral noise and differ from the line profiles of HR~7672~B's spectra.

These variations may be attributed to the fixed temperature and pressure values used in the model, which correspond to conditions at the summit of Mauna Kea.
Ideally, the T-P profile should be set for each molecule.
In the Earth's atmosphere, \ce{H2O} is concentrated at lower altitudes (0 -- 10 km), while \ce{O2} is distributed across a broader altitude range (0 -- 120 km) \citep[e.g., Figure 11 of ][]{10.1051/0004-6361/201833282}.
Therefore, introducing parameters for the T-P profiles specific to each molecule could address the observed deviations.

\subsection{Clouds}
\label{sec:model_cloud}
In the cool atmospheres of brown dwarfs, certain gaseous molecules undergo condensation, transforming into liquid or solid cloud particles.
The predominant effect of these clouds is an increase in opacity.
According to the predictions by equilibrium thermochemical models, a variety of refractory materials condense in L dwarfs, such as corundum (\ce{Al2O3}), iron (\ce{Fe}), enstatite (\ce{MgSiO3}), and forsterite (\ce{Mg2SiO4}), and form cloud layers \citep{10.1086/310356, 10.1006/icar.2001.6740, 10.1007/3-540-30313-8_1, 10.1086/375492, 10.1088/0004-637X/716/2/1060}.
When observing wavelengths with low gas opacity and the absence of clouds, we observe that the flux comes from hotter, deeper atmospheric layers.
The presence of clouds restricts the depth from which L dwarfs can radiate, thereby decreasing the flux.

For our study, we adopted a simplified cloud model characterized by two parameters: the pressure at the cloud layer's top ($P_\mathrm{top}$) and the opacity ($\tau_{\mathrm{cloud}}$).
This relationship is given by: 
\begin{equation}
  \label{eq:cloud}
  \tau^{\prime}(\nu) = \tau(\nu) + \tau_{\mathrm{cloud}}, \quad (P > P_\mathrm{top}),
\end{equation}
where $\tau(\nu)$ and $\tau^{\prime}(\nu)$ is an original and updated opacity, respectively.
In this paper, we fixed the value of $\tau_{\mathrm{cloud}}$ to be $500$, assuming the optically thick cloud is placed at the pressure $P_\mathrm{top}$, so that it is significantly larger than other opacity sources.
Although this assumption does not care much about details such as cloud species or opacity gradient, our primary objective in this paper is to search for evidence of cloudy conditions in HR~7672~B's atmosphere by this model.

\subsection{Effects of Light Leakage from the Host Star}
\label{sec:leak}

\subsubsection{Search for Contaminations from the Host Star}
\begin{figure*}[]
  \begin{center}
  \gridline{\fig{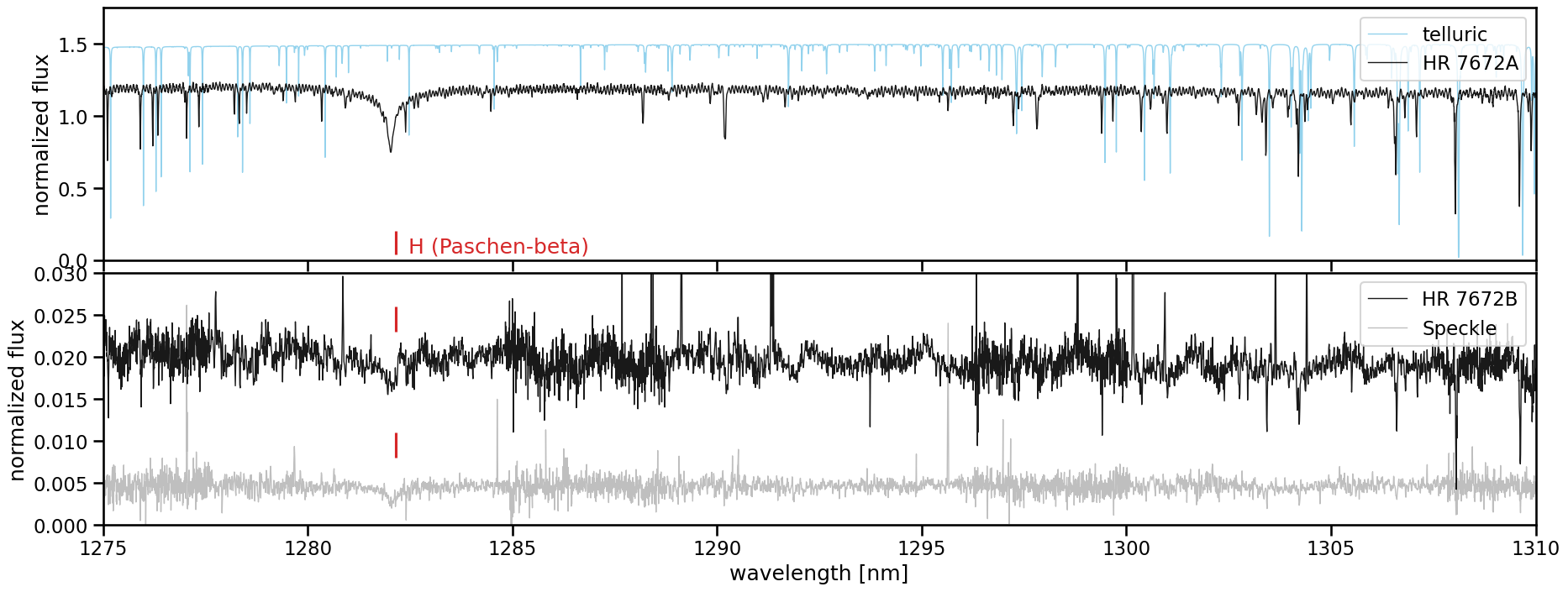}{0.8\linewidth}{(a) around the \ce{H} line (Paschen-$\beta$) at 1282 nm}}
  \gridline{\fig{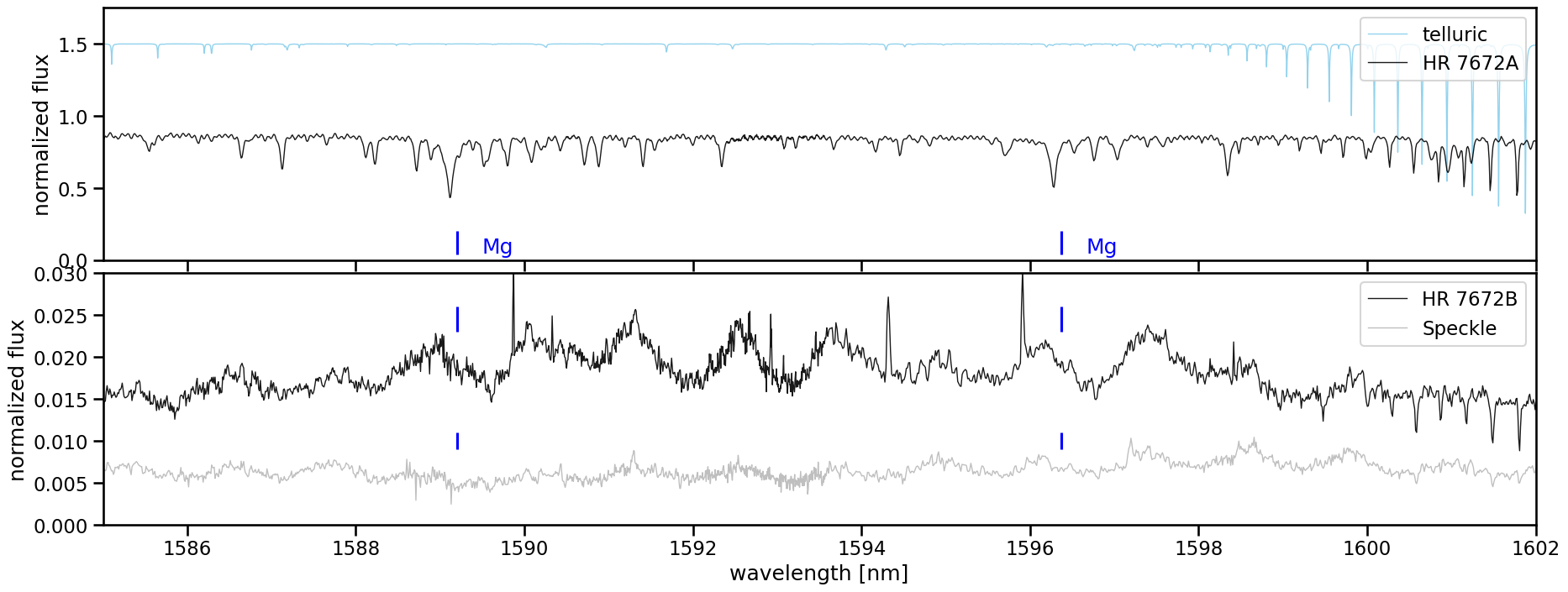}{0.8\linewidth}{(b) around absorption lines of Mg at 1589 nm}}
  \gridline{\fig{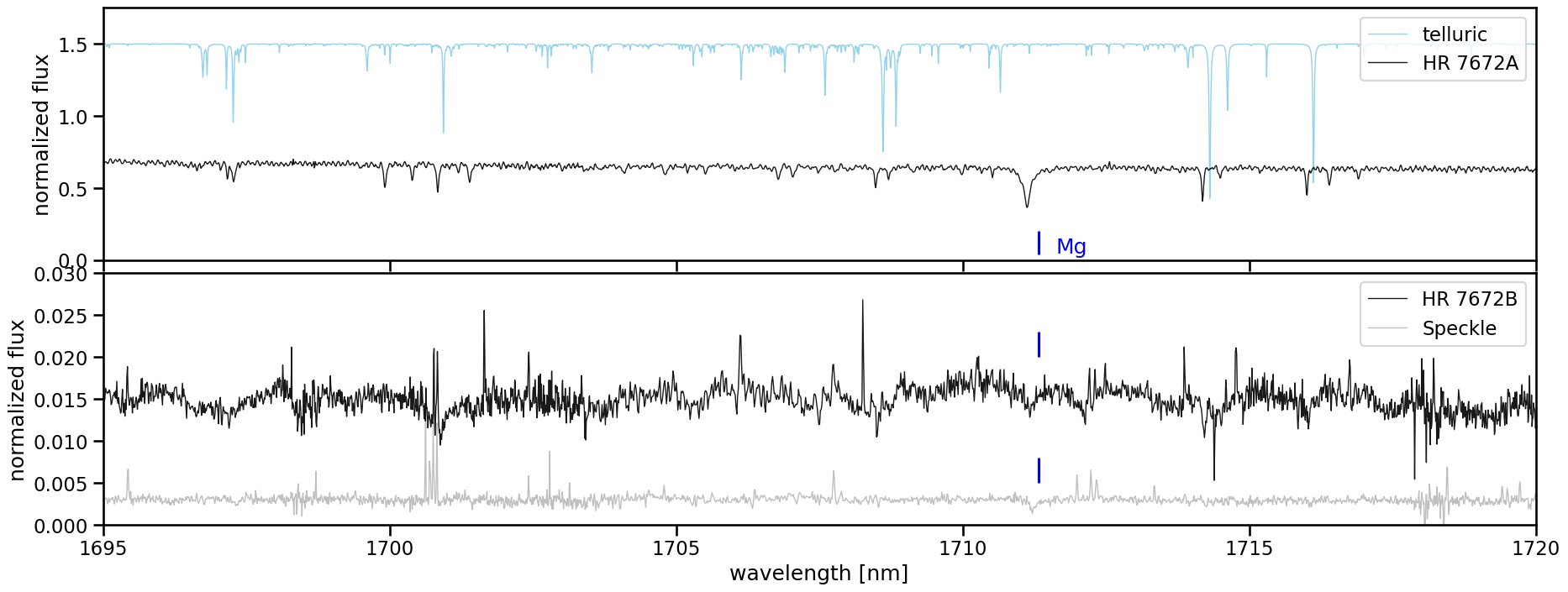}{0.8\linewidth}{(c) around absorption lines of Mg at 1711 nm}}
  \end{center}
  \caption{Comparison of the spectra of HR~7672~A and B was performed to assess light leakage from the host star in HR~7672~B's spectrum. Panels (a)–(c) display the spectra around the strong absorption lines from the host star, as indicated by vertical colored lines, following the wavelengths referenced in \citet{10.1088/0067-0049/185/2/289}. Note that only the barycentric velocity correction for each observation date was applied; the radial velocity correction was not applied, so the wavelengths in the figure are slightly shifted from the expected line positions.}
  \label{fig:HR7672A_HPaMg}
\end{figure*}

We investigated the potential contamination of the HR~7672~B spectra by light from its host star, HR~7672~A.
To assess this, we specifically examined the wavelength regions corresponding to strong absorption features of the host star, using the IRTF Spectral Library \citep{10.1088/0067-0049/185/2/289} as a reference.
Notably, the hydrogen absorption (Paschen-$\beta$) at 1282 nm in the J band of the host star was discernible in both the science and speckle halo fiber spectra, as shown in Figure \ref{fig:HR7672A_HPaMg} (a). 
On the other hand, the presence of other spectral lines, such as \ce{Mg} lines at the H band, was less pronounced due to the contamination of the long-period noise or lower S/N of the host star's absorption lines than the Paschen-$\beta$ line, as shown in Figures \ref{fig:HR7672A_HPaMg} (b) and (c). 
Importantly, absorption lines in the HR~7672~A spectrum are sharper than those in HR~7672~B's due to their rotational velocities are significantly different ($v \sin i \sim 4 \, \mathrm{km\,s^{-1}}$ for HR~7672~A; \citet{2023AJ....165..164B}, and $v \sin i \sim 41\, \mathrm{km\,s^{-1}}$ for HR~7672~B; this work).
For those reasons, we concluded that the contamination of individual absorption lines is negligible for our analysis. 
However, as indicated by the presence of the Paschen-$\beta$ and \ce{Mg} lines in the target spectrum, there should exist light contamination from the host star, which likely influences the line depths in the normalized spectrum of HR~7672~B. 

\subsubsection{Evaluation of the Level of the Contamination around Paschen-$\beta$}
\label{sec:leak_evaluate}

We estimated the level of contamination in HR~7672~B's spectrum using the following model, which combines the scaled host star's spectrum and the brown dwarf model:
\begin{align}
  \label{eq:scale_star}
  F_\mathrm{model} (\nu) &= \left( x_\mathrm{A} F_\mathrm{A}(\nu) + (1-x_\mathrm{A}) F_\mathrm{B, model} (\nu) \right) \cdot T_\mathrm{B}(\nu) \notag \\
  &= \left( x_\mathrm{A} \frac{F_\mathrm{A, obs}(\nu)}{T_\mathrm{A}(\nu)} + (1-x_\mathrm{A}) F_\mathrm{B, model} (\nu) \right) \cdot T_\mathrm{B}(\nu), 
\end{align}
where $x_\mathrm{A}$ is the scaling coefficient, $F_\mathrm{model} (\nu)$, $F_\mathrm{A, obs} (\nu)$, and  $F_\mathrm{B, model} (\nu)$ represent the model spectrum, the observed host star's spectrum, and the modeled companion's spectrum, respectively.
Both $F_\mathrm{A,obs} (\nu)$ and $F_\mathrm{B,model} (\nu)$ are normalized spectra.
The terms $T_\mathrm{A}(\nu)$ and $T_\mathrm{B}(\nu)$ represent the telluric transmittance models at the time of observation for the host star and the companion, respectively.

In this study, we assume that $T_\mathrm{A}(\nu)$ exactly matches $T_\mathrm{B}(\nu)$, which simplifies Equation \eqref{eq:scale_star} to
\begin{equation}
  \label{eq:scale_star_reduced}
  F_\mathrm{model} (\nu) = x_\mathrm{A} F_\mathrm{A,obs}(\nu) + (1-x_\mathrm{A}) F_\mathrm{B, model} (\nu) T_\mathrm{B}(\nu).
\end{equation}
Although this assumption does not strictly hold, as HR~7672~A and B were observed on different dates, we adopt it because the airmass and PWV values were similar, and the differences around the telluric lines in their spectra remained small. 
This suggests that the impact of differing observing conditions is minor and that our model parameterization is sufficiently robust to compensate for these differences.
For reference, the airmasses ($\mathrm{AMs}$) at the time of observation were 1.148 for HR~7672~A and 1.024 for HR~7672~B.
If an airmass correction were applied to the host star's spectrum, it would be performed via $F_\mathrm{A, corrected} = F^{\mathrm{AM_B}/\mathrm{AM_A}}_\mathrm{A, obs}=F^{0.89}_\mathrm{A, obs}$.
The Precipitable Water Vapour (PWV) values on the observation dates of A and B were 2.17 mm and 1.89 mm, respectively, based on measurements from the JCMT water vapour monitor (WVM) \citep{2001ApJ...553.1036W, 2013MNRAS.430.2534D}, assuming comparable atmospheric conditions at Subaru owing to their co-location on Mauna Kea.
Differences in barycentric velocity between observations and in radial velocity between HR~7672~A and B were also not corrected for, as their effects are negligible.
The former is $4.9\,\mathrm{km\,s^{-1}}$, corresponding to a wavelength shift of $\sim$0.2 \AA at $1.5\,\mu \mathrm{m}$, which is comparable to the spectral sampling interval of $\sim0.15$ \AA at the same wavelength.
The latter is estimated to be less than $2\,\mathrm{m\,s^{-1}}$ based on the orbital fitting by  \citet{10.3847/1538-3881/ab04a8}.

Using Equation \eqref{eq:scale_star_reduced}, we estimated the level of contamination focused only on the region around the Paschen-$\beta$ line, as it was difficult to assess the region around the \ce{Mg} absorption at 1711 nm due to the low S/N.
Specifically for this evaluation: (i) corrections for airmass and barycentric velocity were applied; (ii) a uniform prior was imposed on the logarithm of the scaling coefficient ($\log{\mathrm{scale\_star}} = \log{x_\mathrm{A}}$) with $\mathcal{U}(-3, 0)$; and (iii) normal priors were adopted for the temperature at a reference pressure ($T_0$) with $\mathcal{N}(\bar{T_0}, 200\; \mathrm{K})$, the mass ($M_\mathrm{p}$) with $\mathcal{N}(72.7, 0.8\; M_\mathrm{Jup})$, and the surface gravity ($\log g$) with $\mathcal{N}(\overline{\log g}, 0.5)$, where barred symbols ($\bar{T_0}\;\mathrm{and}\;\overline{\log g}$) represent the mean values from the chi-square-based optimization. 
These normal priors were introduced to prevent poor fitting due to the limited number of absorption lines, particularly of \ce{H2O}, in the brown dwarf spectrum within this short wavelength range. 
From this retrieval, the contamination of Paschen-$\beta$ from the speckle halo in the HR~7672~B's spectrum is estimated to be $65 \pm 3\,\%$ of the HR~7672~B's spectrum.

To account for this in our retrievals, a parameter for the amount of the leakage of host star light in the J band was fixed to $63\,\%$ ($= -0.2$ in common logarithm).
We think that the deviation from the true value can be compensated by introducing a parameter $a$, which adjusts for the linear trend of the spectrum. 

In the H band, there are no strong absorption lines as Paschen-$\beta$, and major absorption lines are hidden by telluric or noise, making direct evaluation difficult. 
By giving the uniform distribution as a prior distribution for the parameter that scales HR~7672~A's spectrum, which is modeled in the same way as Equation \eqref{eq:scale_star}, it should be possible to probabilistically evaluate the posterior distribution from Bayesian estimation.

\subsection{Remarks on Other Parameters}
\label{sec:model_others}

We introduced free parameters to correct for discrepancies between the observed high-resolution spectrum and the model, which likely arise from reduction-related effects.
Such discrepancies have also been addressed in previous studies, either by applying a constant offset \citep[e.g., ][]{10.3847/1538-3881/ac56e2} or by correcting wavelength-dependent variations using the blackbody spectrum \citep{2025ApJ...988...53K}.
The latter approach accounts for distortion introduced when normalizing spectra with blaze functions extracted from a FLAT spectrum.
The shape of the blaze function reflects the blackbody radiation of the flat lamp, but its temperature is inherently uncertain due to factors such as lamp aging.
\citet{2025ApJ...988...53K} reported a best-fit blackbody temperature of 1234 K for the flat lamp of IRD/Subaru, based on a standard star observed in their study.

In contrast to \citet{2025ApJ...988...53K}, no such standard star was observed on the same night in our case. 
Therefore, we adopted the following polynomial form as a practical alternative to blackbody-based calibration.
In our implementation, we approximated the correction function as a first-order polynomial in wavenumber $\nu$.
This function is normalized at a reference wavenumber $\nu_0$, consistent with the normalization applied to both the observed and modeled spectra during the retrieval process (Section~\ref{sec:exojax_model}).
The corrected model spectrum, as used in Equation~\eqref{eq:scale_star_reduced}, is thus expressed as:
\begin{equation}
    F_\mathrm{B,model}(\nu) = \frac{1}{a + b (\nu - \nu_0)}\hat{F}_\mathrm{B,model}(\nu),
\end{equation}
where $\nu_0$ represents the mean wavenumber of the reference order, $a$ and $b$ are constant values treated as free parameters, and $\hat{F}_\mathrm{B,model}$ is the modeled companion's spectrum normalized by the median flux within a narrow window centered at $\nu_0$.
The constant term $a$ accounts for deviations when approximating the blackbody function with a polynomial, and is constrained by a Gaussian prior centered at 1.0.
Because the wavelength dependence of the correction may vary between the J and H bands due to the temperature dependence of blackbody spectra, we introduced separate correction parameters for each band.

The J-band magnitude fitting described in Section \ref{sec:model_opa} was computed by scaling the flux of the spectral model by a factor of $(R_\mathrm{p}/d)^2$, where $R_\mathrm{p}$ is the radius and $d$ is the distance.
The distance was fixed at $d = 17.71$ pc in our analysis. 
This value is based on the \textit{Gaia} DR2 parallax reported by \citet{2018yCat.1347....0B}, with an uncertainty of $\pm 0.02$ pc. 
Small differences in the adopted distance have only a minor effect on spectral scaling and do not affect the retrieval results. 
The radius was implicitly determined based on the mass and surface gravity, with both parameters treated as retrieval parameters.
In this paper, two types of prior distributions were applied for the mass parameter: in the ``mass-constrained retrieval,'' the dynamical mass measurement of $M=72.7 \pm 0.8\mathrm{M_{Jup}}$ from \citet{10.3847/1538-3881/ab04a8} was adopted as prior information, and in the ``mass-free retrieval,'' a uniform prior distribution of 1--150 $\mathrm{M_{Jup}}$ was used.
The H-band magnitude calculation was not included in this analysis, because the wavelength range used in this paper does not cover the absorptions of key molecules like \ce{CH4}, a crucial opacity source in the H band.

The Earth's barycentric motion was corrected after the retrieval analysis by using {\sf radial\_velocity\_correction} function from {\sf astropy}.
The RV values shown in Figures \ref{fig:corner_JH_cloud} and \ref{fig:corner_JH_wocloud}, as well as those reported in Table \ref{table:parameters_results}, are barycentric-corrected.

\subsection{Model Likelihood and HMC Settings}
\label{sec:likelihood}

The retrieval process is carried out by modeling the likelihood, assuming that the observational noise for both the high-resolution spectrum $\sigma_\mathrm{p}$ and J-band magnitude $\sigma_\mathrm{J}$ obeys an independent normal distribution,
\begin{align}
    \label{eq:likelihood}
    \mathcal{L} =& \frac{1}{\sqrt{2\pi\sigma_\mathrm{J}^2}} \exp\left( -\frac{(D_\mathrm{J}-M_\mathrm{J})^2}{2\sigma_\mathrm{J}^2} \right) \notag \\
    & \times \prod_i \frac{1}{\sqrt{2\pi(\sigma_{\mathrm{p},i}^2 + \sigma_j^2)}} \exp\left( -\frac{(D_i-F_{\mathrm{model},i})^2}{2(\sigma_{\mathrm{p},i}^2 + \sigma_j^2)} \right),
\end{align}
where $D_\mathrm{J}\pm\sigma_\mathrm{J}=14.39\pm0.20$ is the observed J-band magnitude by \citet{10.1051/0004-6361:20031216}, $M_\mathrm{J}$ is the modeled J-band magnitude, $D_i\pm\sigma_{\mathrm{p},i}$ is the observed high-resolution spectrum and its photon noise, $F_\mathrm{model}$ is the model spectrum.
To account for potential underestimation of the noise, we introduce a jitter term ($\sigma_j$), representing additional noise sources beyond photon noise, and treat it as a free parameter in the retrieval.
The posterior distributions were sampled using the Hamiltonian Monte Carlo (HMC) method with No-U-Turn Sampler \citep[NUTS; ][]{10.48550/arXiv.1111.4246}, using {\sf NumPyro} \citep{10.48550/arXiv.1912.11554}.

We ran MCMC with the HMC-NUTS algorithm and ignored the first 1000 samples as warmup, then derived the results from 1000 samples.
The computation was performed on the GPU of NVIDIA A6000 (48 GB), taking $\sim$10 days in total for each setup. 
The convergence diagnostic values, $\hat{R}$, for all the parameters were below 1.1 in all retrievals, indicating that the simulations converged successfully \citep{2014bda..book.....G}.

All parameter ranges are based on the results of an independent optimization, performed separately from the main analysis.
We searched for the parameter set that minimizes the chi-squared statistic, defined as
\begin{align}
    \chi^2 &= \chi^2_{\mathrm{phot,\, J}} + \chi^2_{\mathrm{spec}} \\
    &= \frac{(D_\mathrm{J}-M_\mathrm{J})^2}{2\sigma_\mathrm{J}^2} + \sum_{i} \frac{(D_i-F_{\mathrm{model},i})^2}{2\sigma_{\mathrm{p},i}^2}
\end{align}
using the Adam optimization algorithm implemented in JAXopt \citep{10.48550/arXiv.2105.15183}.

\begin{table*}[ht]
\centering
\begin{threeparttable}
\caption{Parameter Settings}
\label{table:parameters_priors}
\begin{small}
\begin{tabular}{lccccc}

  \hline\hline
  Parameter & Symbol & Unit & Prior Type & Lower & Upper \\
  & & & & or Mean & or Std \\
  \hline

  Temperature at 1 bar (for Clear Sky) & $T_0^{1\mathrm{bar}}$ & K & Uniform & 600 & 2600 \\
  Temperature at cloud top (for With Clouds) & $T_0^{P\mathrm{top}}$ & K & Uniform & 1200 & 3200 \\
  Power-law index & alpha & -- & Uniform & 0 & 1 \\
  Surface gravity & $\log g$ & cgs for $g$ & Uniform & 4 & 6.5 \\
  Mass (for Mass-Constrained) & $M_\mathrm{p}$ & $M_\mathrm{Jup}$ & Normal & 72.7 & 0.8 \\
  Mass (for Mass-Free) & $M_\mathrm{p}$ & $M_\mathrm{Jup}$ & Uniform & 1 & 150 \\
  Radial velocity & RV & $\mathrm{km\,s^{-1}}$ & Uniform & -30 & -10 \\
  Projected rotation speed & v$\sin i$ & $\mathrm{km\,s^{-1}}$ & Uniform & 30 & 60 \\
  \ce{H2O} Volume Mixing Ratio\tnote{a} & log\ce{H2O} & -- & Uniform & -15 & 0 \\
  \ce{FeH} Volume Mixing Ratio\tnote{a} & log\ce{FeH} & -- & Uniform & -15 & 0 \\
  Jitter noise & sigma & -- & Exponential & 10 & -- \\
  \hline
  \textit{Telluric} \\
  Wavelength shift of telluric lines & $v_\mathrm{tel}$ & $\mathrm{km\,s^{-1}}$ & Uniform & -0.5 & 0.5 \\
  Coefficient of \ce{H2O} opacity & logbeta\_\ce{H2O} & -- & Uniform & 15 & 25 \\
  Coefficient of \ce{CO2} opacity & logbeta\_\ce{CO2} & -- & Uniform & 15 & 25 \\
  Coefficient of \ce{CH4} opacity & logbeta\_\ce{CH4} & -- & Uniform & 15 & 25 \\
  Coefficient of \ce{O2} opacity & logbeta\_\ce{O2} & -- & Uniform & 15 & 25 \\
  \hline
  \textit{Clouds} \\
  Pressure at the top of clouds & $\log P_\mathrm{top}$ & bar for $P_\mathrm{top}$ & Uniform & -1 & 3 \\
  Opacity of clouds & taucloud & -- & Fixed & 500 & -- \\
  \hline
  \textit{Host star's leakage light} \\
  Fraction of the host star's light (J band) & logscale\_star\_y & -- & Fixed & -0.2 & -- \\
  Fraction of the host star's light (H band) &   logscale\_star\_h & -- & Uniform & -3 & 0 \\
  \hline
  \textit{Deviation from normalized spectrum} \\
  ($a+b(\nu-\nu_0)$, $\nu$: wavenumber [cm$^{-1}$]) & & & & & \\ 
  Flux offset & a & -- & Normal & 1 & 0.1 \\
  Slope & b & $10^{-2}$ cm$^{-1}$ & Uniform & -0.1 & 0.1 \\
  \hline
\end{tabular}
\end{small}
\vspace{0.5em}
\begin{tablenotes}
\footnotesize
\item[a] Read Section \ref{sec:model_opa} for the used molecular species in each wavelength range.
\end{tablenotes}
\end{threeparttable}
\end{table*}

\section{Results}
\label{sec:exojax_results}

In this section, we present the retrieval results from four different setups:
In Section \ref{sec:exojax_results_massconstrained}, we used the dynamical mass measurement of $M=72.7 \pm 0.8\mathrm{M_{Jup}}$ for the mass prior, while in Section \ref{sec:exojax_results_massfree}, we adopt a uniform prior ranging from 1 to 150 $M_{\mathrm{J}}$.
For each mass prior, we also examine models both including and excluding cloud opacity.

The wavelength ranges used were where long-period noise is not significant, corresponding to the spectrum in the J band (orders 43 -- 45, $1.265 < \lambda < 1.3 \mu \mathrm{m}$) and a shorter wavelength range in the H band (orders 57 -- 60, $1.48 < \lambda < 1.54 \mu \mathrm{m}$).
Although short-period noise presents within this wavelength range, its effects are negligible for our analysis because the absorption features of HR~7672~B are much broader (lower frequency) due to its rapid rotation.

\subsection{Mass-Constrained Retrieval}
\label{sec:exojax_results_massconstrained}
\subsubsection{Results from a Cloudy Model (Mass-Constrained)}
\label{sec:exojax_results_feducial}

\begin{figure*}[]
    \begin{center}
     \includegraphics[keepaspectratio, width=\linewidth]{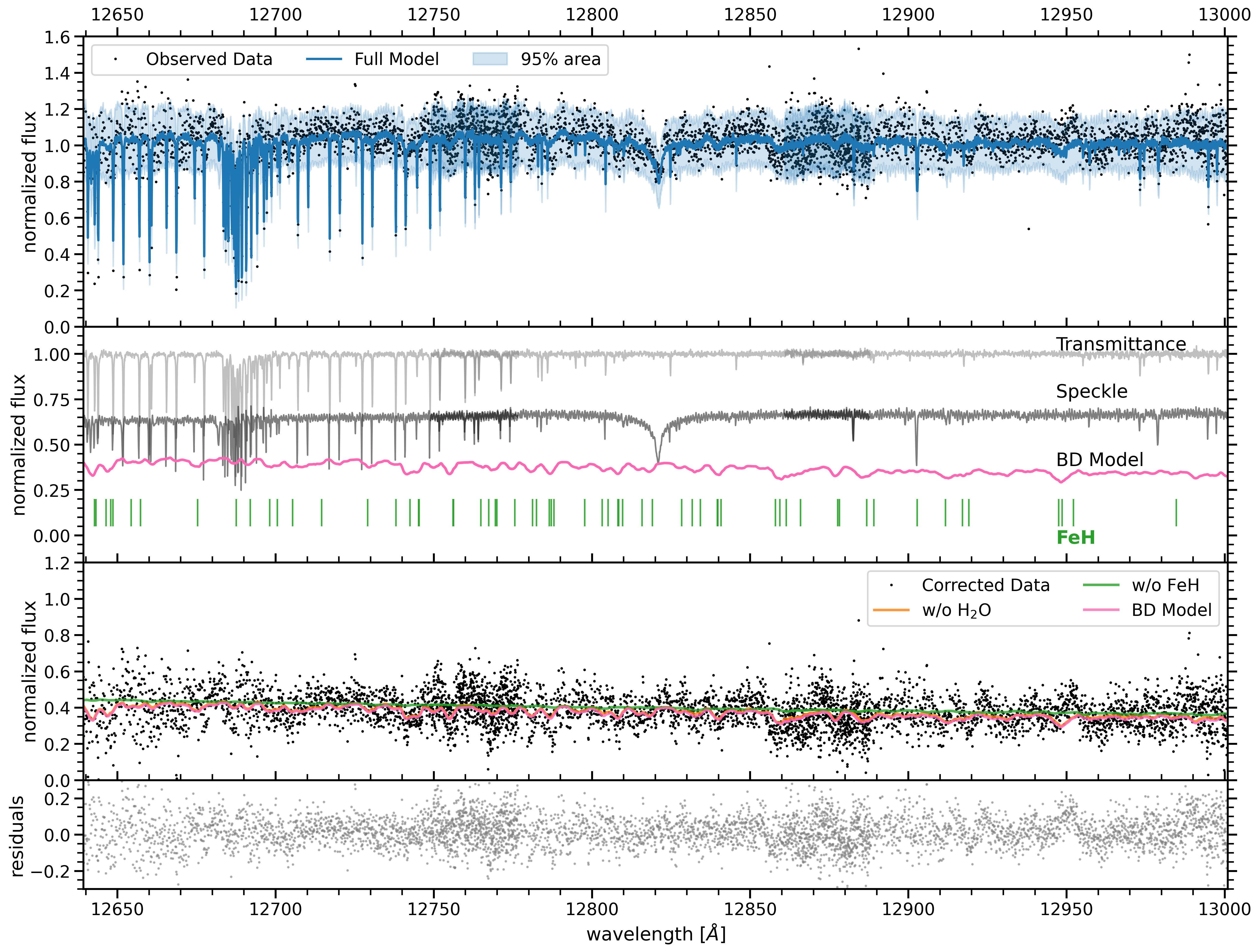}
  \end{center}
  \caption{Retrieval results for the J-band spectrum with the model including cloud opacity. The top panel is the observed spectrum of HR~7672~B and the full model with 95 \% credible intervals. The second panel shows the leakage of the host star's light (dark gray), that is the scaled spectrum of the observed host star, HR~7672~A, retrieved telluric transmittance (light gray), and the brown dwarf's spectrum model (pink). Prominent \ce{FeH} line positions are indicated by vertical green lines. The third panel shows the spectrum of HR~7672~B, which was corrected by dividing by the telluric lines and subtracting the scaled HR~7672~A's spectrum (black dots). The pink lines in the third panel also represent the retrieved brown dwarf model. Other colored lines represent models excluding specific molecules; specifically, the orange lines correspond to models without \ce{H2O}, and the green lines to models without \ce{FeH}. The bottom panel shows residuals.}
  \label{fig:hmc_speckle_cloud_y}
\end{figure*}
\begin{figure*}[]
    \begin{center}
     \includegraphics[keepaspectratio, width=\linewidth]{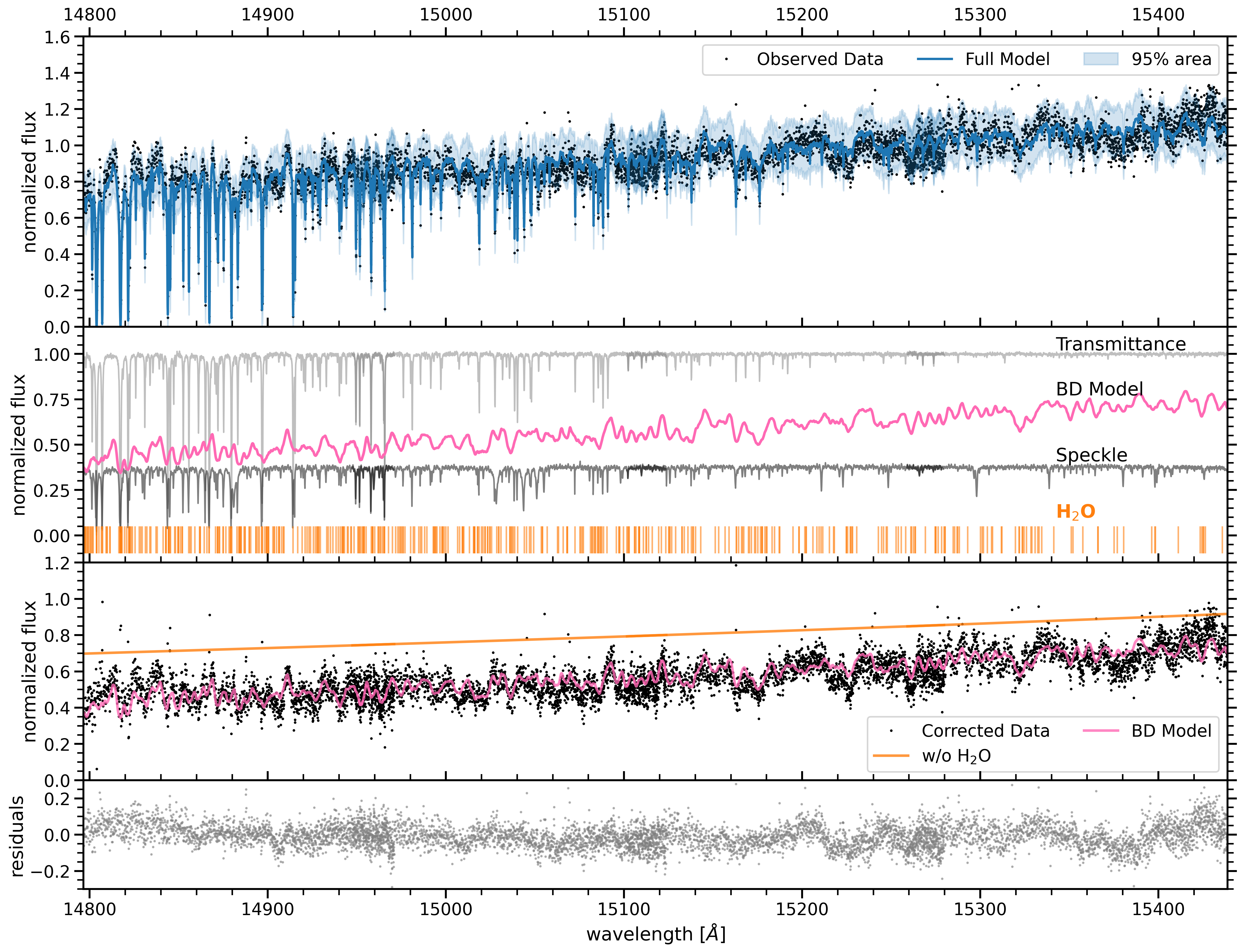}
  \end{center}
  \caption{Same as Figure \ref{fig:hmc_speckle_cloud_y}, but for the H-band spectrum. Prominent \ce{H2O} line positions are indicated by orange lines in the second panel.}
  \label{fig:hmc_speckle_cloud_h}
\end{figure*}

\begin{figure*}[]
  \begin{center}
      \includegraphics[keepaspectratio, width=\linewidth]{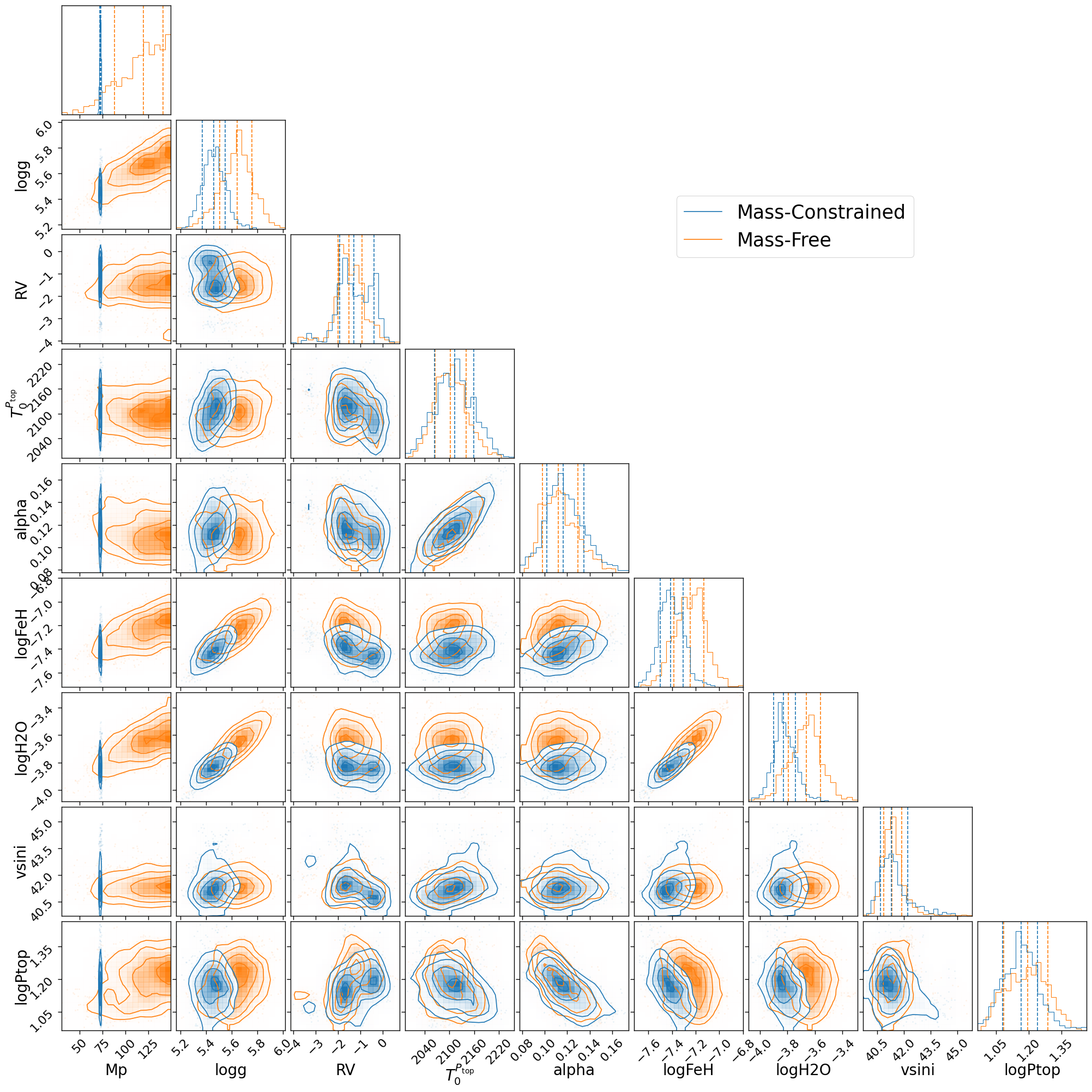}
  \end{center}
  \caption{Posterior distributions of the retrieval for the J- and H-band spectra when a model includes cloud opacity. Only the main parameters are shown here, excluding telluric-related ones.}
  \label{fig:corner_JH_cloud}
\end{figure*} 

The retrieval results with a model that includes cloud opacity are shown in Figures \ref{fig:hmc_speckle_cloud_y} and \ref{fig:hmc_speckle_cloud_h} for the J and H bands, respectively.
The posterior distributions of the main parameters can be found in Figure \ref{fig:corner_JH_cloud}, and the derived parameters are summarized in Table \ref{table:parameters_results}.

Our analysis identified \ce{H2O} and \ce{FeH} as the primary absorbers in the high-resolution spectrum.
The colored vertical lines in the second panels of Figures \ref{fig:hmc_speckle_cloud_y} and \ref{fig:hmc_speckle_cloud_h} indicate prominent line positions of \ce{FeH} and \ce{H2O}, selected based on line strengths ($S_{ij}$) at $T_\mathrm{top}=2112$ K: specifically, $S_{ij} \geq 3\times10^{-19}$ cm for \ce{FeH} and $S_{ij} \geq 5\times10^{-23}$ cm for \ce{H2O}.
In the H band spectrum shown in the third panel of Figure \ref{fig:hmc_speckle_cloud_h}, the absorption features are well represented by the pink line (the brown dwarf's spectrum model), but disappear in the orange line, which was calculated by removing \ce{H2O} from the model (i.e., including only CIA and cloud opacities).
This suggests that \ce{H2O} is a primary source of line opacity for these absorption features.
Similarly, in the J-band spectrum shown in the third panel of Figure \ref{fig:hmc_speckle_cloud_y}, \ce{FeH} (green) lines predominantly appear and the pink line almost overlaps the orange line calculated by removing \ce{H2O} from the model, which means the contribution from \ce{H2O} is small in this wavelength range.

The prediction of the J-band magnitude is 14.48, with a 95 \% credible interval of [13.86, 14.98]. 
This range includes the observed J-band magnitude of $14.39 \pm 0.20$. 

The estimated leakage of the host star's light in the H band, treated as a free parameter, is around 40 \%. 
This is lower than the J band (60 \%), as determined from a spectrum around Paschen-$\beta$. 
Such a result aligns with expectations given the higher fiber coupling efficiency of an SMF on REACH in the H band.

\begin{figure}[]
  \begin{center}
      \includegraphics[keepaspectratio, width=0.9\linewidth]{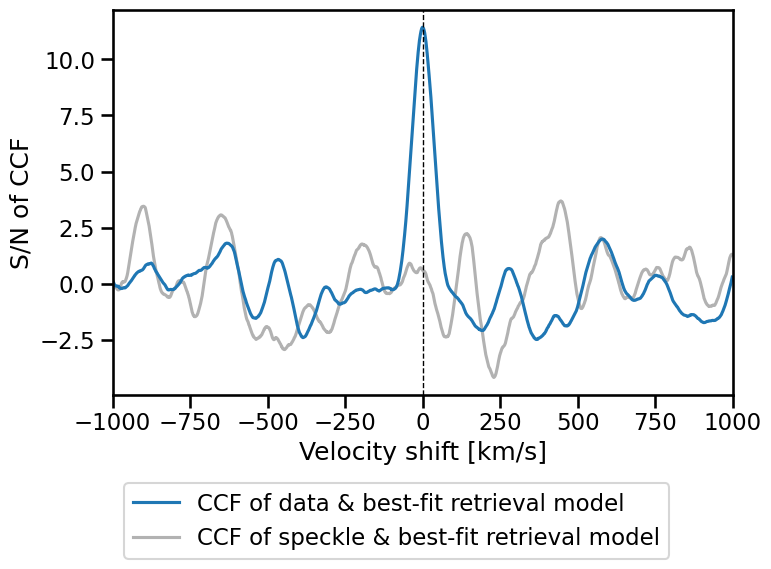}
  \end{center}
  \caption{Cross-correlation function (CCF) of the observed spectrum with the best-fit model (blue line). The model was Doppler-shifted by the retrieved RV, such that the CCF peak is expected to appear at $0~\mathrm{km\,s^{-1}}$ (vertical dashed line). For comparison, the grey line shows the CCF of the speckle halo spectrum, simultaneously obtained from the speckle halo fiber during the HR 7672 B observation.}
  \label{fig:ccf_bestfit}
\end{figure}

To validate the detection of brown dwarf signal, we performed a least-squares cross-correlation function (CCF) analysis, following the method of \citet{2021AJ....162..148W}, also used by \citet{2024ApJ...971....9H}; for detailed methodology, refer to \citet{2021AJ....162..148W}.
Briefly, this method estimates the maximum-likelihood values of the companion and speckle fluxes as a function of RV shift.
We used companion spectrum and transmittance templates derived from the best-fit mass-constrained cloudy model, and adopted the observed HR 7672 A spectrum as the stellar speckle template.
The companion template was Doppler-shifted from $-1000$ to $+1000~\mathrm{km\,s^{-1}}$ in $1~\mathrm{km\,s^{-1}}$ steps.
CCFs are expressed in S/N, normalized by the standard deviation of the CCF wings ($-1000$ to $-500$ and $+500$ to $+1000~\mathrm{km\,s^{-1}}$), assumed to be signal-free.
Figure~\ref{fig:ccf_bestfit} (a) shows a clear CCF peak at $0~\mathrm{km\,s^{-1}}$ with $\mathrm{S/N} = 11.4$, while no significant peak appears for the speckle halo spectrum.
The CCF analysis supporting the detection of each molecular species is shown in Appendix \ref{sec:ccf_mols}.

\begin{figure}[]
  \begin{center}
      \includegraphics[keepaspectratio, width=\linewidth]{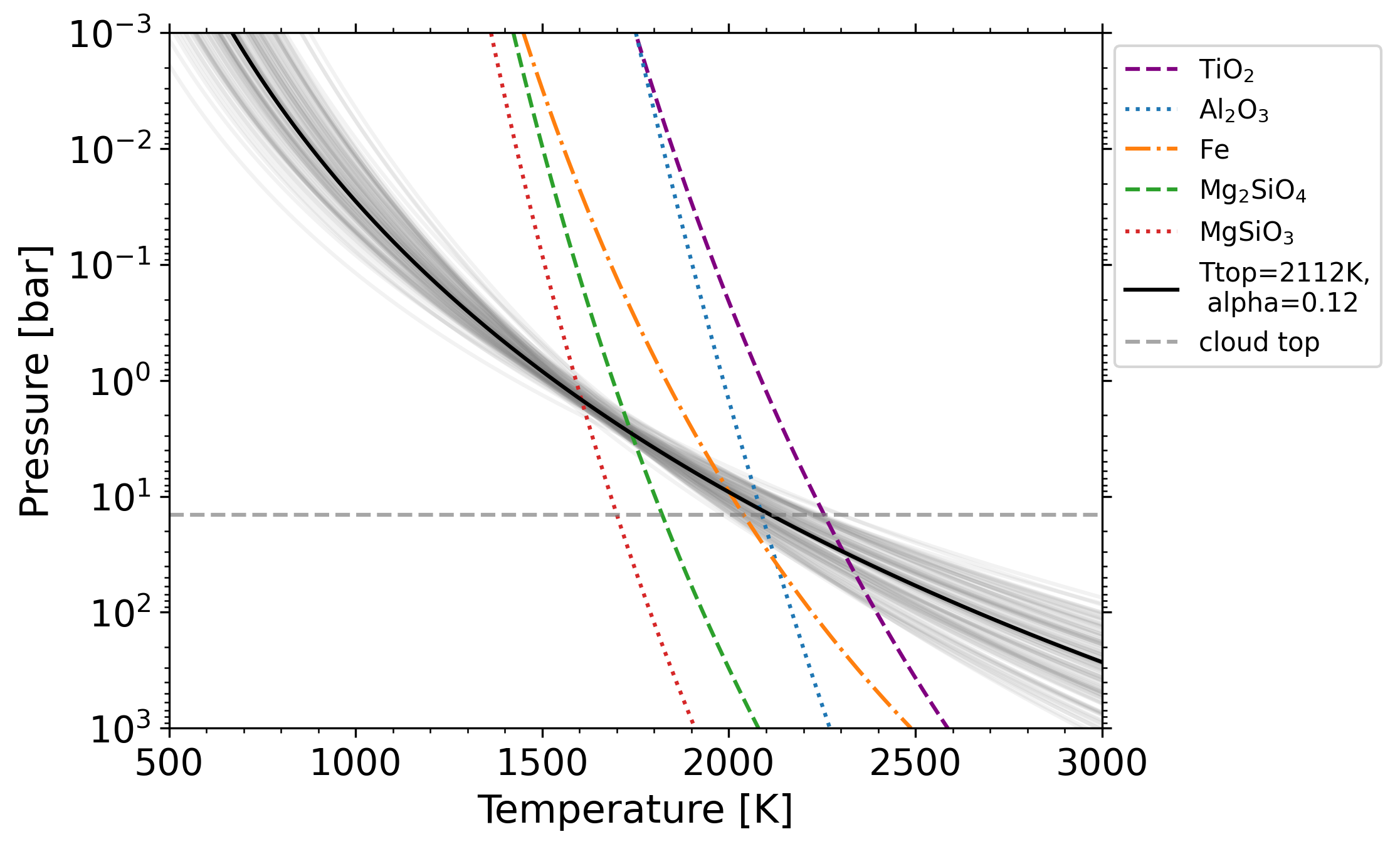}
  \end{center}
  \caption{Comparison of the retrieved T-P profile (black line) with condensation curves (colored dashed lines). The gray lines represent 100 random samples from the posterior distributions, while the gray dashed line indicates the retrieved pressure at the cloud top. References for the condensation curves: \citet{10.1051/0004-6361:20010937} for \ce{TiO2}; \citet{2017MNRAS.464.4247W} for \ce{Al2O3}; \citet{10.1088/0004-637X/716/2/1060} for \ce{Fe}, \ce{Mg2SiO4}, and \ce{MgSiO3}. The metallicity is set to be $Z=-0.04$ according to \citet{10.3847/1538-3881/ac56e2}, except for \ce{TiO2}, as its treatment of metallicity in \citet{10.1051/0004-6361:20010937} is uncertain.}
  \label{fig:T-P_cloud_wTP}
\end{figure}

In this analysis, the presence of clouds was preferred, even though the model also allowed for a clear sky condition by setting the cloud top pressure ($P_\mathrm{top}$) to a sufficiently high value.
Consequently, we proceeded to examine the parameters derived from the retrieval using this model to ensure the accuracy and reliability of our findings.
First, to check the types of clouds that might condense under the derived atmospheric conditions, we have overlaid the retrieved T-P profile on various condensation curves, as shown in Figure \ref{fig:T-P_cloud_wTP}.
In this figure, the best-fit T-P profile is represented by the black line, while the grey lines are generated using 100 random samples from the posterior distribution.
Given that the temperature at the top of the cloud is approximately 2112 K, which lies between the condensation temperatures of \ce{TiO2}, \ce{Al2O3}, and \ce{Fe} and above the temperature where \ce{Mg2SiO4} remains gaseous, it is plausible that clouds composed of \ce{TiO2}, \ce{Al2O3}, or \ce{Fe} could form in this atmosphere and contribute to the near-infrared opacity observed. 

\begin{figure}[]
  \begin{center}
      \includegraphics[keepaspectratio, width=\linewidth]{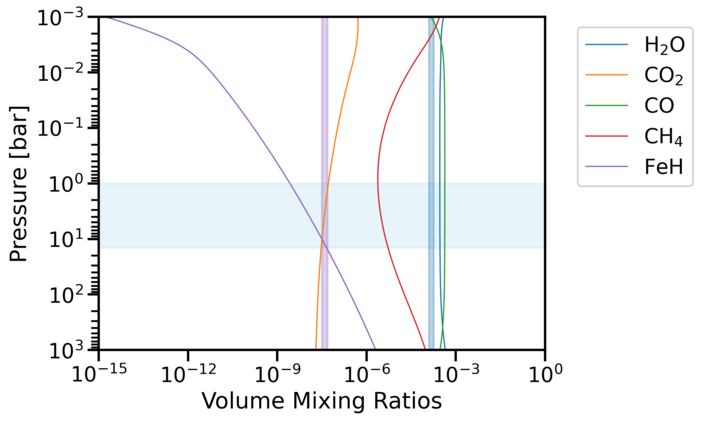}
  \end{center}
  \caption{Abundances assuming thermochemical equilibrium (solid lines). We set $\mathrm{C/O}=0.56$ and $\mathrm{[Fe/H]}=-0.04$ for this calculation. The retrieved $1\sigma$ values of VMRs for \ce{H2O} and \ce{FeH} are shown by vertical shaded regions. According to the contribution functions, we observe the atmosphere in the region around $10^{0}<P<10^{1.1}$ bar, shown as the horizontally shaded region.}
  \label{fig:fastchem_JH_cloud}
\end{figure}

Then, to verify the plausibility of the abundances retrieved in our study, we checked the thermochemical equilibrium corresponding to the derived T-P profile.
For this, we utilized {\sf FastChem} \citep{10.1093/mnras/sty1531} to calculate the chemical composition under thermochemical equilibrium conditions.
The values for C/O and [Fe/H] were set to stellar values ($\mathrm{C/O}=0.56$ and $\mathrm{[Fe/H]}=-0.04$), as reported by \citet{10.3847/1538-3881/ac56e2}.
Figure \ref{fig:fastchem_JH_cloud} shows the results using the median T-P profile, with the retrieved value and the $1\sigma$ range for \ce{H2O} and \ce{FeH} shown as shaded areas.
The horizontally blue-shaded region in this figure corresponds to the observing atmospheric pressure range, approximately $10^{0}<P<10^{1.1}$ bar.
This range is determined by examining the contribution function, which indicates the altitudes at which spectral features originate in the atmosphere. 
At $P_\mathrm{top}=1.16$, the retrieved VMR for \ce{FeH} is almost consistent with the thermochemical equilibrium prediction (1.2$\sigma$ deviation), while that for \ce{H2O} is underestimated (3.3$\sigma$ deviation).
Whether these deviations reflect true underestimations or are due to model limitations or limited spectral coverage remains uncertain. 
Future observations with broader wavelength coverage will be essential to confirm their origin.

\subsubsection{Results from a Clear Sky Model (Mass-Constrained)}
\label{sec:exojax_results_clearsky}
\begin{figure*}[]
  \begin{center}
      \includegraphics[keepaspectratio, width=\linewidth]{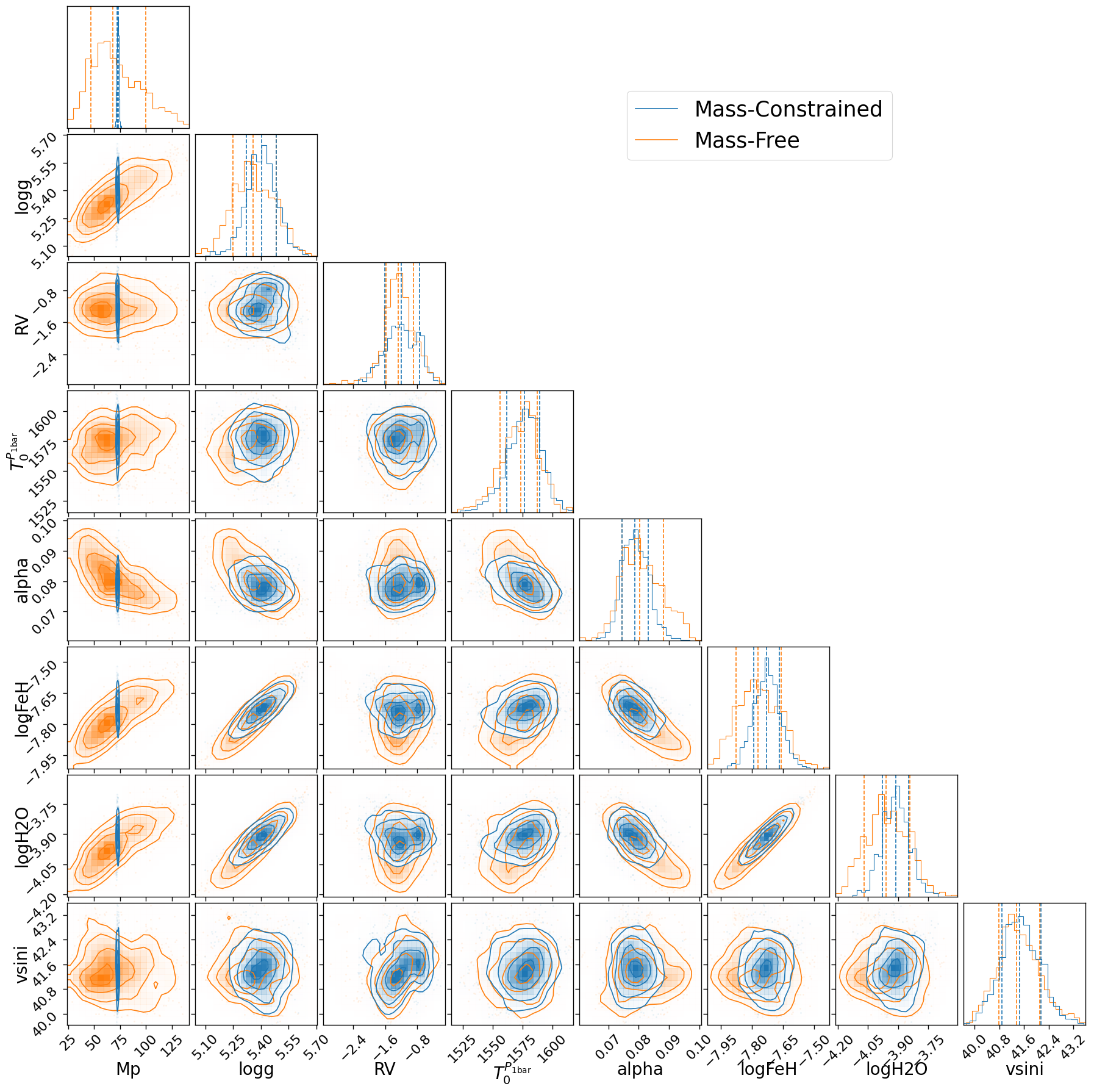}
  \end{center}
  \caption{Posterior distributions of the retrieval for the J- and H-band spectra with a clear sky model. Only the main parameters are shown here, excluding telluric-related ones.}
  \label{fig:corner_JH_wocloud}
\end{figure*}

To explore whether cloud opacity is essential for explaining the observed high-resolution spectrum, we conducted a retrieval using a clear sky model that does not include cloud opacity.
The posterior distributions of the main parameters can be found in Figure \ref{fig:corner_JH_wocloud}. 

In this case, we were also able to reproduce both the high-resolution spectrum and the J-band magnitude, with almost the same accuracy.
The prediction of the J-band magnitude is 14.40, with a 95 \% credible interval of [13.85, 14.93].
This range includes the observed J-band magnitude of $14.39 \pm 0.20$. 

The reason why this cloudless condition can explain the observed spectrum was explored through its contribution function. 
We confirmed that CIA opacity primarily contributed to the spectral continuum in this case, whereas cloud opacity was responsible for the continuum in the cloudy model. 
Since CIA opacity is effective in the lower atmosphere, the T-P profile was adjusted to be steeper with a smaller value of $\alpha$, as listed in Table \ref{table:parameters_results}, to bring the temperature at the altitude around $\tau=1$ closer to that of the cloudy model.

\subsubsection{Implications of Continuum Opacity in the Mass-Constrained Retrieval}
To summarize the results from the mass-constrained retrieval, our analysis suggests that an additional continuum opacity source, beyond line opacity, is required to reproduce the observed spectrum in this wavelength region. 
When cloud opacity is included in the model, it serves as the continuum opacity. 
However, even in the absence of cloud opacity, CIA opacity effectively plays this role to match the observed spectrum.
Since the prior range of cloud top pressure includes values where cloud opacity does not affect the spectrum, the results from the cloudy model (Section \ref{sec:exojax_results_feducial}) suggest that the cloudy condition is preferred. 
Therefore, we consider the result with the model that includes cloud opacity to be the fiducial case.
These behaviors will be further discussed in Section \ref{sec:exojax_discussion_opa}.

To further quantify the model comparison, we computed the difference in the Bayesian Information Criterion ($\Delta \mathrm{BIC}$) between the retrieval settings, using the following definition: 
\begin{equation}
    \mathrm{BIC} = k \ln N -2 \ln \hat{\mathcal{L}}
\end{equation}
where $k$ is the number of free parameters estimated by the model ($k=20$ for the cloudy model and $k=19$ for the clear sky model), $N$ is the number of data points, and $\hat{\mathcal{L}}$ is the maximum of the likelihood function (Equation \eqref{eq:likelihood}).
When selecting among multiple models, the one with the lowest BIC is considered preferred.
Using the mass-constrained retrieval with a cloudy model as the reference, the mass-constrained retrieval with a clear sky model yields $\Delta \mathrm{BIC}=21.9$.
According to the threshold guidelines proposed by \citet{Kass1995}, a $\Delta \mathrm{BIC}$ greater than 10 constitutes very strong evidence against the model with the higher BIC, indicating that the cloudy model is preferred.

\subsection{Mass-Free Retrieval}
\label{sec:exojax_results_massfree}

The motivation for performing mass-free retrieval is to enable mass estimation without relying on evolutionary models. 
This approach is especially important for long-period planetary-mass companions and isolated brown dwarfs, for which dynamical mass measurements are often unavailable. 
In such cases, evolutionary models are typically used, but these suffer from large uncertainties due to poorly constrained ages and metallicities.
As discussed in \citet{2025ApJ...988...53K}, high-resolution spectra can provide an alternative method of mass estimation that does not depend on evolutionary models.
By applying mass-free retrieval here, we aim to further explore and validate this method for broader applications.

\subsubsection{Results from a Cloudy Model (Mass-Free)}
\label{sec:exojax_results_uniformmass_cloud}

The posterior distributions from the mass-free retrieval with a cloudy model are overplotted in Figure \ref{fig:corner_JH_cloud}. 
All parameters, except for mass, converged to nearly the same values as those from the mass-constrained retrieval. 
Specifically, we observed that (i) the pressure at the cloud top converged to a value nearly identical to that in the mass-constrained retrieval, and (ii) the upper limit of the mass was unconstrained.

Point (i) supports the mass-constrained result, indicating that the cloudy atmosphere is preferred for the observed spectrum.
This cloudy condition likely influences the mechanism underlying Point (ii), which will be discussed in detail in Section \ref{sec:exojax_discussion_gravity}.

Before moving to the detailed discussion, we first clarify that the determination of mass is equivalent to the determination of surface gravity in this retrieval analysis, which uses both the high-resolution spectrum and the J-band magnitude.
The radius can be estimated from the observed broadband magnitude using the relation $L \propto R^2 T^4$, where the observed intensity ratio of the absorption lines provides information about the temperature. 
Thus, since surface gravity is calculated from $g = GM/R^2$, the ratio $M/g$ is constrained. 

This $M/g$ constraint is evident from the linear trend in the $M_\mathrm{p}$ and $\log g$ distribution panel shown in Figure \ref{fig:corner_JH_cloud}, and the radius can be derived as $R_\mathrm{p} = 0.82^{+0.10}_{-0.07} \; R_\mathrm{Jup}$.
While $\log g$ appears to have converged, this is due to the upper limit on the mass being set to $150M_\mathrm{Jup}$.  

Therefore, comparing the result with the mass-constrained retrieval in Section \ref{sec:exojax_results_feducial}, with a uniform mass prior, the model that includes cloud opacity cannot constrain the upper limit of surface gravity.

\subsubsection{Results from a Clear Sky Model (Mass-Free)}
\label{sec:exojax_results_uniformmass_clear}

The posterior distributions resulting from the mass-free retrieval with a clear sky model are overplotted in Figure \ref{fig:corner_JH_wocloud}. 
The derived parameters are largely consistent with those from the mass-constrained retrieval, except for larger credible intervals, particularly for the mass.
This model also yields $\Delta \mathrm{BIC} = 21.2$ compared to the mass-constrained retrieval with a cloudy model, indicating a strong preference for the latter.

In contrast to the result with a cloudy model in the previous section (\ref{sec:exojax_results_uniformmass_cloud}), both mass and gravity were constrained in this case.
This is because CIA opacity contributes to the spectral continuum in the observed wavelength range, which can constrain surface gravity through the following mechanism, as also discussed in \citet{2025ApJ...988...53K}.
When CIA opacity is absent and only line opacity contributes to the spectrum, a degeneracy arises between surface gravity ($g$) and the mixing ratio of molecules. 
This is because the line opacity is expressed as:
\begin{equation}
    \label{eq:opaline}
    d\tau_{\mathrm{line},i} = -\frac{x_i \sigma_i}{\mu m_u g}dP \propto \frac{x_i}{g},
\end{equation}
where $x_i$ and $\sigma_i$ are the volume mixing ratio and the cross-section of species $i$, respectively, and $\mu$ and $m_u$ are the mean molecular weight and atomic mass unit, respectively.
However, CIA opacity helps resolve this degeneracy, as it is inversely proportional to gravity:
\begin{equation}
    \label{eq:opacia}
    d\tau_{\mathrm{CIA},i,j} = -\beta_{i,j} n_i n_j \frac{k_\mathrm{B}T}{\mu m_u g} \frac{dP}{P} \propto \frac{1}{g},
\end{equation}
where $\beta_{i,j}$ is the CIA absorption coefficient, and $k_\mathrm{B}$ is the Boltzmann constant.
Therefore, since this model does not include cloud opacity, which would otherwise obscure the wavelength range where CIA opacity works as a continuum, the degeneracy associated with gravity was resolved.

\begin{table*}[ht]
\centering
\begin{threeparttable}
\caption{Summary of Retrieval Results}
\label{table:parameters_results}
\begin{small}
\begin{tabular}{lccccc}
  \hline \hline
    & Unit & \multicolumn{2}{c}{\textbf{Mass-Constrained}} & \multicolumn{2}{c}{Mass-Free} \\
    & & \textbf{With Clouds}\tnote{a} & Clear sky & With Clouds & Clear sky \\ \hline
  $T_0^{1\mathrm{bar}}$ & K & -- & $1577 ^{+13} _{-15}$ & -- & $1574 ^{+14} _{-17}$ \\
  $T_0^{P\mathrm{top}}$ & K & $2112 ^{+46} _{-49}$ & -- &$2102 ^{+39} _{-37}$ & -- \\
  alpha & -- & $0.12 ^{+0.02} _{-0.01}$ & $0.08 ^{+0.00} _{-0.00}$ & $0.11 ^{+0.02} _{-0.01}$ & $0.08 ^{+0.01} _{-0.01}$ \\
  log$g$ & cgs for $g$ & $5.46 ^{+0.09} _{-0.09}$ & $5.40 ^{+0.08} _{-0.08}$ & $5.64 ^{+0.11} _{-0.14}$ & $5.36 ^{+0.13} _{-0.11}$ \\
  $M_\mathrm{p}$ & $M_\mathrm{Jup} $& $72.71 ^{+0.81} _{-0.85}$ & $72.73 ^{+0.83} _{-0.88}$ & $>30.6$ & $67.87 ^{+31.81} _{-20.79}$ \\
  RV & $\mathrm{km\,s^{-1}}$ & $-1.29 ^{+0.90} _{-0.63}$ & $-1.19 ^{+0.45} _{-0.42}$ & $-1.53 ^{+0.60} _{-0.47}$ & $-1.27 ^{+0.37} _{-0.31}$ \\
  vsini & $\mathrm{km\,s^{-1}}$ & $41.28 ^{+0.92} _{-0.61}$ & $41.46 ^{+0.68} _{-0.57}$ & $41.31 ^{+0.57} _{-0.44}$ & $41.35 ^{+0.76} _{-0.57}$ \\
  log\ce{H2O} & -- & $-3.83 ^{+0.09} _{-0.07}$ & $-3.91 ^{+0.06} _{-0.07}$ & $-3.66 ^{+0.10} _{-0.13}$ & $-3.96 ^{+0.12} _{-0.11}$ \\
  log\ce{FeH} & -- & $-7.41 ^{+0.10} _{-0.09}$ & $-7.73 ^{+0.06} _{-0.06}$ & $-7.24 ^{+0.11} _{-0.14}$ & $-7.77 ^{+0.11} _{-0.11}$ \\
  sigma & -- & $0.06 ^{+0.00} _{-0.00}$ & $0.06 ^{+0.00} _{-0.00}$ & $0.06 ^{+0.00} _{-0.00}$ & $0.06 ^{+0.00} _{-0.00}$ \\
  \hline
  \textit{Telluric} \\
  $v_\mathrm{tel}$ & $\mathrm{km\,s^{-1}}$ & $-0.04 ^{+0.08} _{-0.09}$ & $0.01 ^{+0.08} _{-0.08}$ & $-0.03 ^{+0.07} _{-0.08}$ & $-0.01 ^{+0.08} _{-0.08}$ \\
  logbeta\_\ce{H2O} & -- & $21.48 ^{+0.02} _{-0.02}$ & $21.51 ^{+0.02} _{-0.02}$ & $21.49 ^{+0.02} _{-0.02}$ & $21.51 ^{+0.02} _{-0.02}$ \\
  logbeta\_\ce{CO2} & -- & $21.28 ^{+0.34} _{-3.28}$ & $21.56 ^{+0.15} _{-0.33}$ & $21.34 ^{+0.29} _{-2.11}$ & $21.44 ^{+0.24} _{-5.26}$ \\
  logbeta\_\ce{CH4} & -- & $17.49 ^{+1.67} _{-1.69}$ & $17.53 ^{+1.65} _{-1.62}$ & $17.57 ^{+1.61} _{-1.70}$ & $17.27 ^{+1.85} _{-1.65}$ \\
  logbeta\_\ce{O2} & -- & $24.20 ^{+0.03} _{-0.03}$ & $24.19 ^{+0.03} _{-0.03}$ & $24.20 ^{+0.02} _{-0.03}$ & $24.20 ^{+0.03} _{-0.03}$ \\
  \hline
  \textit{Clouds} \\
  logPtop & bar for $P_\mathrm{top}$ &$1.16 ^{+0.08} _{-0.08}$ & -- & $1.20 ^{+0.09} _{-0.11}$ & -- \\
  \hline
  \textit{Host star's leakage light} \\
  logscale\_star\_h & -- &$-0.43 ^{+0.02} _{-0.02}$ & $-0.46 ^{+0.02} _{-0.02}$ & $-0.43 ^{+0.02} _{-0.02}$ & $-0.46 ^{+0.02} _{-0.02}$ \\
  \hline
  \textit{Corrections} \\
  a\_y\tnote{b} & -- & $1.15 ^{+0.50} _{-0.42}$ & $1.18 ^{+0.43} _{-0.41}$ & $1.20 ^{+0.47} _{-0.44}$ & $1.16 ^{+0.42} _{-0.44}$ \\
  b\_y\tnote{b} & $10^{-2}$ cm$^{-1}$ & $-0.13 ^{+0.05} _{-0.06}$ & $-0.12 ^{+0.04} _{-0.05}$ & $-0.14 ^{+0.05} _{-0.06}$ & $-0.12 ^{+0.05} _{-0.04}$ \\
  a\_h & -- & $1.27 ^{+0.45} _{-0.46}$ & $1.22 ^{+0.44} _{-0.40}$ & $1.21 ^{+0.45} _{-0.43}$ & $1.23 ^{+0.49} _{-0.40}$ \\
  b\_h & $10^{-2}$ cm$^{-1}$ & $0.11 ^{+0.04} _{-0.04}$ & $0.10 ^{+0.04} _{-0.03}$ & $0.10 ^{+0.04} _{-0.04}$ & $0.11 ^{+0.04} _{-0.04}$ \\
  \hline \hline
  \textit{Derived values} \\
  $T$ at 1 bar & K & $1544 ^{+26} _{-31}$ & -- & $1544 ^{+24} _{-25}$ & -- \\
  $R_\mathrm{p}$ & $R_\mathrm{Jup}$ & $0.81 ^{+0.09} _{-0.08}$ & $0.86 ^{+0.08} _{-0.07}$ & $0.82 ^{+0.10} _{-0.07}$ & $0.88 ^{+0.09} _{-0.10}$\\
  $\Delta \mathrm{BIC}$ & -- & -- & 21.9 & N/A\tnote{c} & 21.2 \\
  \hline
\end{tabular}
\end{small}
\vspace{0.5em}
\begin{tablenotes}
\footnotesize
\item Note --- The presented errors correspond to the $1\sigma$ values.
\item[a] The fiducial result.
\item[b] The suffixes are set to '\_y', but they are for the J band.
\item[c] $M_\mathrm{p}$ is unconstrained, and the maximum log-likelihood is unreliable.
\end{tablenotes}
\end{threeparttable}
\end{table*}

\section{Discussion}
\label{sec:exojax_discussion}

\subsection{Cloud Opacity and CIA: Roles in Spectral Continuum Modeling}
\label{sec:exojax_discussion_opa}

\begin{figure*}[]
  \gridline{\fig{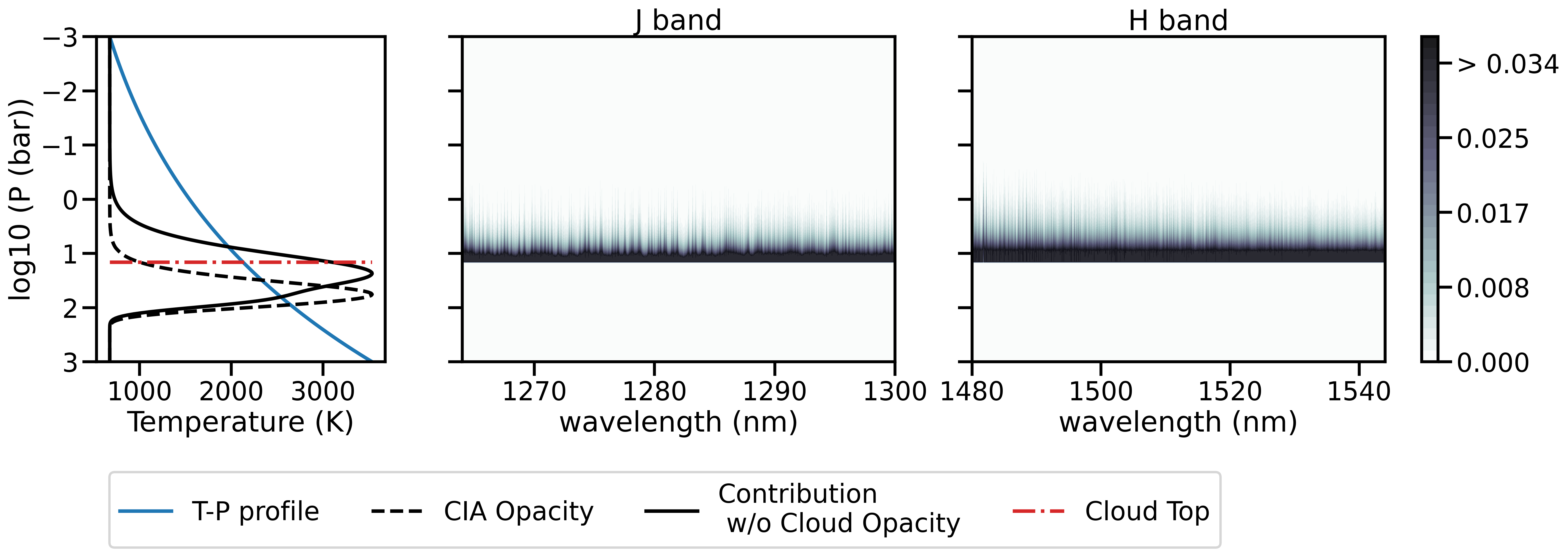}{0.9\linewidth}{(a) w/ Clouds \& Mass-Constrained}}
  \gridline{\fig{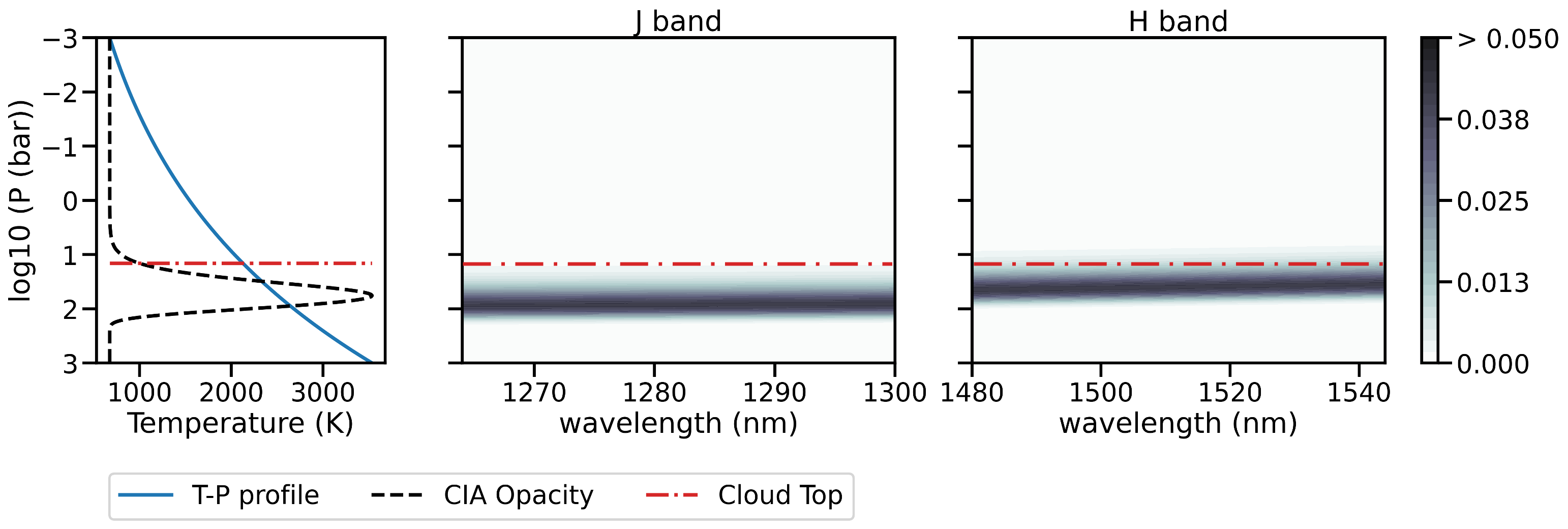}{0.9\linewidth}{(b) only CIA}}
  \caption{Spectrally averaged contributions overlaid with T-P profiles (left panel) and contribution functions (right panel). (a) Results for the mass-constrained cloudy model, showing the total opacity from the retrieval. (b) Results from (a), but isolating only the CIA contribution, excluding cloud opacity. The plotted wavelength range spans from the start of the J-band spectrum to the end of the H-band spectrum used in our retrieval analysis. The white background between the J and H bands corresponds to the wavelength gap not covered in our retrievals.}
  \label{fig:cf}
\end{figure*} 

We found that the key factor distinguishing the retrieval results between the cloudy model and the clear sky model was the opacity source responsible for the spectral continuum. 
Figure \ref{fig:cf} (a) shows the contribution function from the mass-constrained retrieval with cloud opacity included.
The contribution function indicates the altitudes at which spectral features originate in the atmosphere, as given by: 
\begin{equation}
    cf(P) = B(\lambda,T) \frac{de^{-\tau}}{d\log(P)},
\end{equation}
where $B(\lambda,T)$ is the Plank function, $\tau$ is the optical depth, and $P$ is the pressure \citep[e.g., ][]{1987tpaa.book.....C, 1998Sci...282.2063G}\footnote{The negative sign may be mistakenly missing in the exponential factor in \citet{1998Sci...282.2063G}.}.
Figure \ref{fig:cf} (a) clearly illustrates that clouds suppress contributions from the lower atmosphere, emphasizing the necessity of the cloud, even though the model allowed for placing the cloud at a lower altitude to avoid affecting the spectrum.
In contrast, Figure \ref{fig:cf} (b) is plotted using the same parameters as (a) but excludes both line opacity and cloud opacity, showing only the contribution from CIA. 
This figure highlights the wavelength dependence of CIA opacity within the wavelength range used for our retrieval. 
Thus, the preference for a cloudy atmospheric condition suggests that the observed spectrum has a relatively wavelength-independent continuum. 

However, a remaining question is whether the cloud can exist at any altitude above the region where CIA opacity is active, which would result in an unconstrained lower limit for the cloud top pressure.
In contrast, the cloud top pressure was found to converge to $P_\mathrm{top}=10^{1.16}$ bar in our retrieval results.

To investigate the reason for this retrieved cloud altitude, the pressure corresponding to the retrieved cloud top is overplotted as a red dashed line in Figure \ref{fig:cf} (b). 
We found that cloud opacity is the dominant source of the spectral continuum in the J band, while CIA opacity slightly contributes above the cloud in the H band. 
Consequently, we concluded that the observed spectral continuum in the H band exhibits wavelength dependence, whereas the J-band spectrum favors a constant continuum.

We also investigated the degeneracy between cloud presence, T-P profiles, and molecular abundances.
Similar spectral features to those produced by the resulting model can be generated by larger volume mixing ratios and clouds at higher altitudes, with T-P profiles adjusted to maintain the same temperature at the cloud-top pressure.
Under these atmospheric conditions, line opacity will operate at higher altitudes with lower pressures, following the relation $\tau_{\mathrm{line},i} \propto x_i P$ (cf. Equation \eqref{eq:opdepth_line}). 
However, in this case, CIA opacity is obscured by cloud opacity across the entire wavelength range.

To check this degeneracy, we performed a retrieval analysis with the initial value of $P_\mathrm{top}$ set to $10^{0}$ bar instead of $10^{2.5}$ bar. 
The former corresponds to an altitude above where CIA opacity is effective, while the latter corresponds to an altitude below it.
If the observed spectrum favors a wavelength-independent continuum across the entire range, the cloud top pressure should converge to a lower value, completely obscuring CIA opacity.
However, even with the changed initial value, the retrieval converged to the same result, with $P_\mathrm{top}=10^{1.1}$ bar. 
Therefore, the wavelength dependence of the spectral continuum in the H band is certainly favored.

\subsection{Degeneracy between Mass and Surface Gravity}
\label{sec:exojax_discussion_gravity}

\begin{figure*}[]
  \centering
  \includegraphics[keepaspectratio, width=0.8\linewidth]{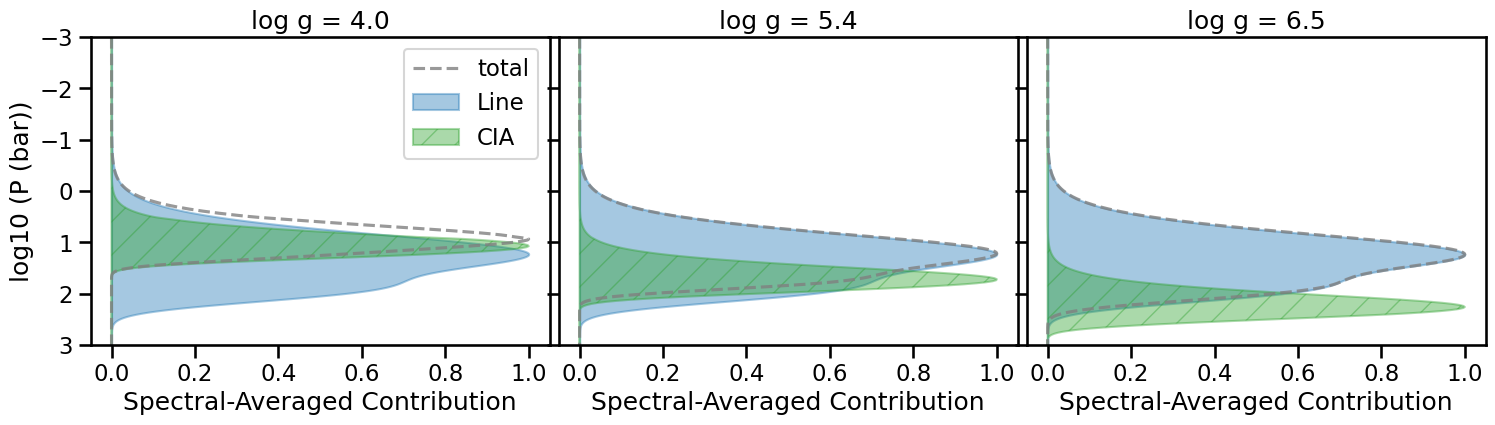}
  \caption{Contribution functions with varying $g$ and $x_i$ while maintaining $g/x_i=\mathrm{const.}$. The atmospheric parameters used to generate all figures are based on the results from the mass-constrained retrieval with a cloudy model. Contributions from both the J- and H-band wavelength ranges used in our retrieval are shown. In the panels from left to right, the surface gravity increases. The blue represents the contribution from line opacity, the green represents that from CIA opacity, and the grey dashed line shows the total opacity from both line and CIA opacity. Note that the case with $\log g = 6.5$ is included, although such a high gravity is not physical for a brown dwarf.}
  \label{fig:cf_peaks}
\end{figure*}

In the mass-free retrieval with a cloudy model (Section \ref{sec:exojax_results_uniformmass_cloud}), the upper limit of the mass was unconstrained, meaning that the upper limit of the surface gravity was also not constrained in the absence of a mass range limit.
As discussed in the previous section (\ref{sec:exojax_discussion_opa}), the cloud must be located at an altitude that reproduces the wavelength dependence of the observed spectrum continuum. 
Additionally, under this condition, the possible T-P profile and molecular abundances are constrained to reproduce the observed spectral features.

However, abundances are still allowed to vary with gravity while maintaining a constant ratio of $g/x_i$.
This is because the atmospheric pressure where line opacity contributes most becomes almost constant if $g/x_i$ is held fixed, shown as blue-shaded regions in Figure \ref{fig:cf_peaks}.
On the other hand, these changes in gravity affect the pressure where CIA opacity contributes most, shown as green shaded regions in Figure \ref{fig:cf_peaks}.
These behaviors are analytically explained for an isothermal atmosphere in Appendix \ref{sec:cf_appendix}. 

This figure suggests that the altitude where each opacity contributes most shifts to the upper atmosphere as gravity decreases, but the dependency on gravity differs between opacity sources.
In particular, when $g$ varies with $x_i$ such that $g/x_i = \mathrm{const.}$, only the pressure where CIA opacity is effective moves to the upper atmosphere at lower gravity, while that of line opacity remains unchanged. 
On the other hand, CIA opacity works in the lower atmosphere when gravity is high, with cloud opacity completely obscuring CIA opacity. 
Thus, the lower limit of gravity is determined by the need to prevent CIA opacity in the upper atmosphere from significantly muting the spectrum. 
In contrast, the upper limit remains unconstrained due to the spectrum’s insensitivity to gravity.

When the mass is well constrained with a strict prior, it also places a constraint on surface gravity via the relation $g \propto M/R^2$, as the possible values of radius are limited by photometric magnitude fitting. 
Therefore, this phenomenon does not occur in the mass-constrained retrieval with a cloudy model.

\subsection{Comparison with Previous Studies}
\label{sec:exojax_discussion_w22}

To date, \citet{10.3847/1538-3881/ac56e2} (hereafter W22) has investigated the atmospheric compositions of HR~7672~B using its high-resolution ($R \sim 35,000$) spectra observed in the K band with KPIC/Keck.
Our retrieval utilized spectra from a different wavelength range, probing partially distinct regions of the atmosphere.
By comparing the contribution functions between their results (Figure 12 in W22) and ours (Figure \ref{fig:cf}), we found that while the K-band observations probe atmospheric pressures of approximately 0.1 to 10 bar, the J- and H-band spectra allow us to examine slightly lower altitudes, corresponding to pressures of approximately 4 to 15 ($\simeq 10^{1.16}$ as $P_\mathrm{top}$) bar.

The retrieval results from W22 suggest a cloudless condition, but we found that our cloudy model was preferred over the clear sky model.
A direct comparison of these results is not possible, as the T-P profile and cloud model differ between the two analyses.
As noted in W22, their cloudless result may be influenced by the known degeneracy, where both clouds and a nearly isothermal T-P profile can lead to shallower absorption lines.
To confirm cloud properties, further analysis combining high-resolution spectra, mid- to low-resolution spectra, and photometry across a broader wavelength range is essential \citep{10.3847/1538-4357/ac8673}.

\subsubsection{Molecular Abundances and $v \sin{i}$}

From the K-band spectrum, W22 successfully retrieved the abundances of \ce{H2O} and \ce{CO}, which were consistent with a quenched condition at a pressure of 10 bar due to atmospheric mixing. 
Furthermore, these detections suggest that the C/O ratio is within $1.5\sigma$ of the abundance of the host star HR7672A.

When comparing the abundance of \ce{H2O}, the only molecular species common to both W22 and our model, the volume mixing ratios and $1\sigma$ errors are $-3.58^{+0.07}_{-0.07}$ and $-3.83^{+0.09}_{-0.07}$, respectively. 
These values are consistent within $3\sigma$. 
Other molecular species, such as \ce{CO} and \ce{CH4}, may contribute to the spectra in the longer wavelength range of the H band \citep{2014PNAS..11112601B, 2018arXiv180408149F}, but we excluded this region from our study due to the presence of long-period noise. 
Future efforts to mitigate periodic noise in the spectra will enable a more detailed comparative analysis of the \ce{C} abundance and C/O ratios between our results and those of W22.

Regarding the projected rotation speed ($v\sin{i}$), W22 reported a value of $45.0 \pm 0.5\, \mathrm{km\,s^{-1}}$. 
In contrast, we derived a value of $41.3^{+0.9}_{-0.6}\, \mathrm{km\,s^{-1}}$, which is consistent with the measurement by \citet{2021JATIS...7c5006D}, who reported $42.6 \pm 0.8\, \mathrm{km\,s^{-1}}$.

\subsubsection{T-P profiles}
\begin{figure}
    \centering
    \includegraphics[width=\linewidth]{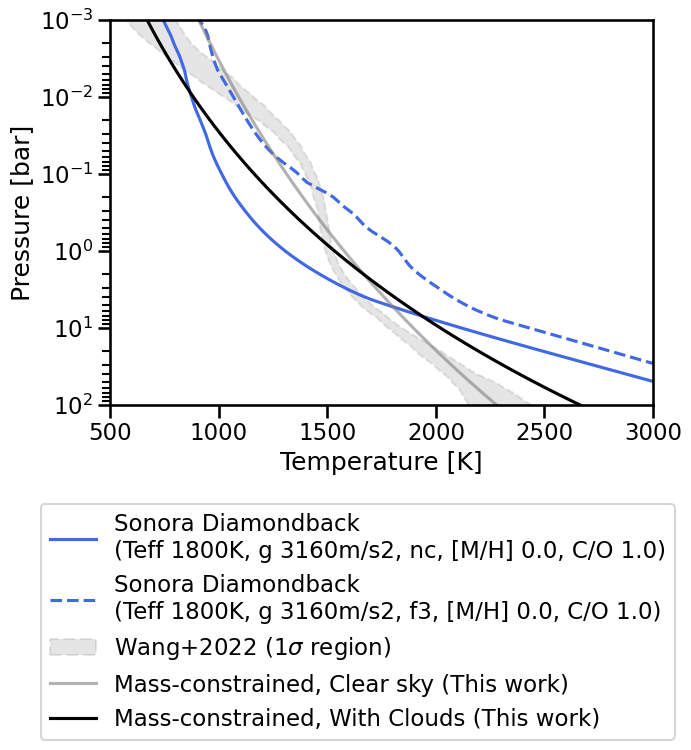}
    \caption{Comparison of T-P profiles. The black and grey lines represent the T-P profiles derived from our retrievals using the cloudy and cloud-free models, respectively. The grey shaded region indicates the 1$\sigma$ range from the retrieval in \citet{10.3847/1538-3881/ac56e2}, reproduced from Figure 12 of that paper. The blue solid and dashed lines show the best-fit cloud-free and cloudy models, respectively, from the Sonora Diamondback grid \citep{2024ApJ...975...59M}, with an effective temperature of 1800 K.}
    \label{fig:tp_compare}
\end{figure}

\begin{figure}[]
  \centering
  \includegraphics[keepaspectratio, width=\linewidth]{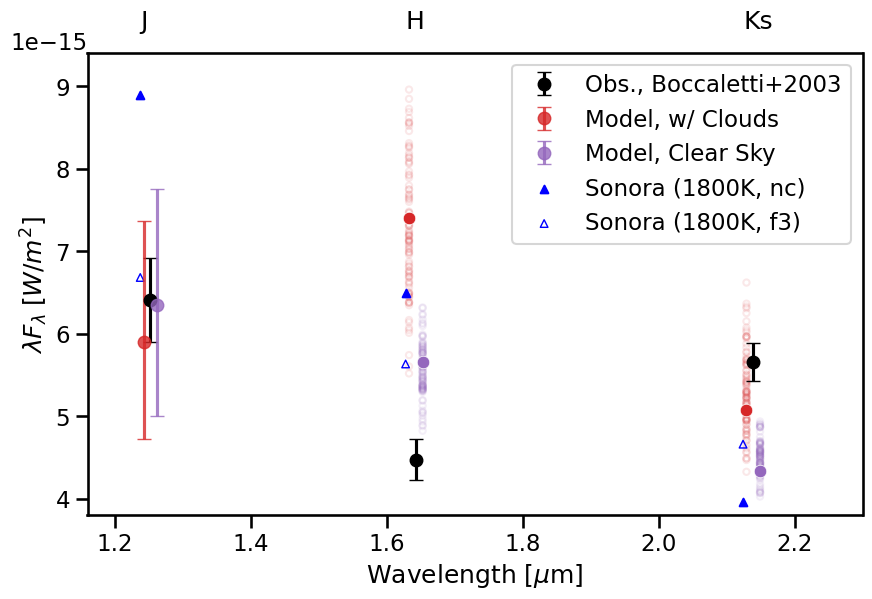}
  \caption{Comparison of fluxes in the J, H, and Ks bands. The black dots represent the observed flux, while the red and purple dots correspond to the fluxes calculated from our model --- red represents the cloudy model and purple represents the clear sky model. The J-band values from our model are derived from predictions and their 95 \% credible intervals. For the H and Ks bands, we plot the values derived using the median parameters of the posterior distribution as filled circles, and 100 random samples from the posterior distributions as translucent circles. The blue triangles represent fluxes computed from the Sonora Diamondback models.}
  \label{fig:flux}
\end{figure}

The T-P profiles are generally consistent between the results presented in W22 and our results from both the cloudy and clear sky models, as shown in Figure \ref{fig:tp_compare}. 
However, a closer examination of the temperature reveals that our cloudy model has a temperature that is approximately 300 K higher around 10 bar, where opacity contributes most in the K band. 
To investigate this discrepancy, we compared the J, H, and Ks-band fluxes with the observed apparent magnitudes from \citet{10.1051/0004-6361:20031216}. 
The results are shown in Figure \ref{fig:flux}, similar to Figure 9 in W22.

The values from our models for the J band are derived from predictions with their 95 \% credible errors detailed in Sections \ref{sec:exojax_results_feducial} and \ref{sec:exojax_results_clearsky}, while the H- and Ks-band values are calculated from forward-modeled spectra using the median values and 100 random samples from the posterior distributions of parameters. 
For this estimation, we also included line opacity from additional molecules, \ce{CO} and \ce{CH4}, using the HITRAN database, with their abundances fixed to the values from the mass-fixed retrieval in W22. 
The filters used are WFCAM/UKIRT for the J and H bands, and NIRC2/Keck for the Ks band.

Since the J-band magnitude is simultaneously fitted in our retrieval analysis, it is consistent for both the cloudy and clear sky models. 
The H-band magnitudes for both models are higher than the observations, a discrepancy also noted in W22. 
Given that both the J- and Ks-band fluxes are consistent with the observed values, this discrepancy is less likely to be explained by the wavelength dependence of clouds, as discussed in W22, since clouds affect shorter wavelengths more than longer wavelengths. 
Instead, this could be due to an underestimation of \ce{CH4} abundance, which is a dominant opacity source around 1.6 $\mu$m. 
Additionally, the absence of \ce{FeH} opacity from the $E^4\Pi$–$A^4\Pi$ band at 1.58 $\mu$m may also contribute to this discrepancy.
This band is not included in the MoLLIST line list from ExoMol but is available in the line list provided by \citet{2010AJ....140..919H}.
However, due to the large uncertainties in the line strengths in this latter source \citep[e.g., ][]{2021ApJS..254...34G}, we chose not to include it in this analysis.
For the Ks-band magnitude, although our results are slightly underestimated, they are closer to the observed values than those shown in Figure 9 of W22 (approximately $4.1 \times 10^{-15} \, \mathrm{[W\,m^{-2}]}$).
Therefore, except for missing opacity in the H band, T-P profiles derived from our retrieval analysis match previous observations.

In Figure \ref{fig:tp_compare}, we also compare the T-P profile derived from the Sonora Diamondback atmospheric models for substellar objects \citep{2024ApJ...975...59M}.
We calculated the J, H, and Ks-band magnitudes from the Sonora model spectra, and searched for the best-fit model to our retrieval result using the cloudy model by minimizing the chi-square value:
\begin{equation}
    \chi^2_{\mathrm{flux}} = \sum_{k=J,H,Ks} \frac{(\lambda F_{\lambda,k}^{\mathrm{retrieved}} - \lambda F_{\lambda,k}^{\mathrm{sonora}})^2}{\sigma_k^2},
\end{equation}
where $\lambda F_{\lambda,k}^{\mathrm{retrieved}}$ and $\sigma_k$ are the flux and its uncertainty derived from our retrieval for $k=$J, H, and Ks bands, as shown by red dots in Figure \ref{fig:flux}, and $\lambda F_{\lambda,k}^{\mathrm{sonora}}$ is the flux calculated from the Sonora model spectra.
To avoid degeneracies in flux due to model parameters, we restricted the search to models with $g=3160\, \mathrm{m\,s}^{-2}$, fixed to a value close to that obtained from our retrieval, and $\mathrm{[M/H]} = 0.0$, assuming that the metallicity of HR~7672~B is similar to that of its host star ($\mathrm{[Fe/H]} = -0.04$; \citet{10.3847/1538-3881/ac56e2}).

We found that the model with an effective temperature of $1800\,\mathrm{K}$, with an uncertainty of $\pm100\,\mathrm{K}$ due to the grid spacing of the Sonora models, and moderately thick clouds characterized by $f_{\mathrm{sed}} = 3$, provided the best fit with $\chi^2_{\mathrm{flux}} = 2.6$.
This result is consistent with our retrieval finding that a cloudy atmosphere is favored.
However, since the cloud treatment in the Sonora models differs from that in our retrieval framework, direct comparisons should be interpreted with caution.
A detailed assessment of the impact of these differences is beyond the scope of this study and is left for future work.
For comparison, when limited to cloud-free models, the best-fit effective temperature remained $1800\pm100\,\mathrm{K}$, but with a higher chi-square value of $\chi^2_{\mathrm{flux}} = 7.6$.
The corresponding T–P profile and fluxes are shown in Figures~\ref{fig:tp_compare} and \ref{fig:flux}.
The temperature around $10\, \mathrm{bar}$ in the best-fit cloud-free Sonora model is broadly consistent with the T–P profile derived from our retrieval, which corresponds to the approximate cloud-top pressure and the region where opacity has the greatest effect on the spectrum. 
The discrepancy in the upper atmosphere suggests that a more flexible T–P profile parameterization may be necessary for retrievals using spectra that cover a broader wavelength range.

\section{Conclusions and Summary}
\label{sec:exojax_conclusion}
In this work, we used the high-resolution spectrum ($R \sim 100,000$) observed by REACH to investigate the atmospheric properties of HR~7672~B.
The framework of the spectrum model was calculated by using \ExoJAX, and we originally developed three additional models; our model can take into account clouds in the brown dwarf's atmosphere and the leakage light of the host star, and simultaneously model telluric absorptions.

We focused on performing a retrieval analysis of the high-resolution spectrum within a part of the J and H bands, along with the photometric magnitude of the J band.
Our analysis successfully identified \ce{H2O} and \ce{FeH} as the primary absorbers in these wavelength ranges, with their retrieved abundances closely aligning with those expected from thermochemical equilibrium.

Both models, whether or not they include cloud opacity, are capable of reproducing the observed spectra. 
This suggests that an additional continuum opacity source beyond line opacity is necessary. 
The retrieval results that include cloud opacity indicate that clouds are required to explain the observed features, as we treated the cloud top pressure as a free parameter.
This result can be explained by the different wavelength dependencies of the continuum opacity sources: CIA opacity and cloud opacity. 
While the pressure where CIA opacity contributes most to the spectrum depends on the wavelength, our cloud model assumes optically thick clouds below a specific altitude, which act as a wavelength-independent continuum.
We found that the inferred cloud altitude corresponds to the pressure where CIA opacity begins to contribute in the H-band spectrum.
Based on this, we conclude that the H-band spectral continuum exhibits wavelength dependence, whereas the J-band continuum remains constant across the examined range.

While this behavior in the spectral continuum may not be exclusively attributed to the presence of clouds, we found that the retrieved T-P profile suggests the potential condensation of molecules such as \ce{TiO2}, \ce{Al2O3}, and \ce{Fe}. 
Future observations, combining medium- and high-resolution spectra over broader wavelength ranges or monitoring time variability, could provide deeper insights into cloud properties and atmospheric dynamics.

This study also represents the first science demonstration of REACH.
We identified previously unknown periodic noise in the faint target spectra observed by REACH and investigated its properties.
The periodic noise significantly affected the longer wavelength range of the H-band spectra, as characterized in Section \ref{sec:fringe}, preventing its use in retrievals. 
Since this noise tends to appear in spectra from both the science fiber and the speckle halo fiber, it may be possible to mitigate it using the Gaussian Process, for example. 
Additionally, our current model has room for improvement, such as incorporating more flexible T-P profiles and advanced cloud modeling.
While our findings demonstrate the potential for detailed atmospheric analysis using a relatively simple model, future updates of these features will enable a more comprehensive understanding of the atmospheric properties of substellar objects.

\begin{acknowledgments}
We are grateful to the anonymous referee for their careful reading and insightful comments, which significantly enhanced the quality of this work.
We thank Hibiki Yama, Hiroyuki Tako Ishikawa, Ji Wang, and Shotaro Tada for their valuable discussions. 
We also extend our gratitude to Yuka Fujii, the chief referee of Y.K.'s thesis defense, which includes the analysis presented in this paper.

This work was partly supported by JST SPRING grant No. JPMJSP2104, JSPS KAKENHI grant Nos. 24K22912 (Y.K.), 21K13984, 22H05150, 23H01224 (Y. Kawashima), 21H04998, 23H00133, 23H01224 (H.K.), and the Graduate University for Advanced Studies.
The development of SCExAO is supported by JSPS KAKENHI grant No. 21H04998.
Numerical computations were in part carried out on the GPU system at the Center for Computational Astrophysics, National Astronomical Observatory of Japan.
This research is based in part on data collected at the Subaru Telescope, which is operated by the National Astronomical Observatory of Japan. We are honored and grateful for the opportunity of observing the Universe from Maunakea, which has the cultural, historical, and natural significance in Hawaii.

The code used in this work is available at GitHub (\url{https://github.com/YuiKasagi/AtmosphericRetrieval_HR7672B}) and archived with Zenodo \citep{kasagi_2025}. 

\end{acknowledgments}

\vspace{5mm}
\facilities{REACH/Subaru Telescope}

\software{numpy \citep{10.1038/s41586-020-2649-2},
          matplotlib \citep{10.1109/MCSE.2007.55},
          scipy \citep{2020NatMe..17..261V},
          astroquery \citep{10.3847/1538-3881/aafc33},
          ExoJAX \citep{10.3847/1538-4365/ac3b4d, 2025ApJ...985..263K},
          PyIRD (Y. Kasagi et al. submitted),
          pandas \citep{reback2020pandas},
          jax \citep{jax2018github},
          astropy \citep{2013A&A...558A..33A,2018AJ....156..123A,2022ApJ...935..167A},
          numpyro \citep{10.48550/arXiv.1912.11554},
          jaxopt \citep{10.48550/arXiv.2105.15183},
          corner \citep{10.21105/joss.00024},
          seaborn \citep{10.21105/joss.03021},
          FastChem \citep{10.1093/mnras/sty1531},
          AtmosphericRetrieval\_HR7672B \citep{kasagi_2025}
          }

\appendix

\section{Detailed properties of periodic noise}
\label{sec:fringe_appendix}

\subsection{Periodic Noise with a Long Period}

Periodic noise, characterized by a period of approximately 1.2 nm, is evident in the spectrum.
While its origin is unknown, it appears more significant in the spectra of faint targets and shows similar patterns for different targets observed on the same night.

\subsubsection{Does Not Appear in the FLAT Spectrum as well as the Outside of an Aperture}

To analyze the periodic nature of the spectra shown in Figure \ref{fig:fringe_spectra}, we employed the Lomb-Scargle periodogram \citep{10.1007/BF00648343, 10.1086/160554}.
The shorter periodicity manifests in the target, FLAT, and speckle spectra, whereas the longer periodicity is evident only in the target and speckle spectra.
The absence of either periodicity in a spectrum with no light, which is extracted by setting apertures shifted 40 pixels from the target light position, affirms that these periodicities are not additive in nature. 
Moreover, in terms of wavelength dependency, the periodicity of this noise gets longer with increasing order, specifically across orders from 63 to 67.

\subsection{Periodic Noise with a Short Period}

Periodic noise, characterized by a period of approximately 0.18 nm ($\sim$ few tens of pixels), with amplitudes ranging from 0.67 \% to 1.3 \%, is also evident in the spectrum.
Given its presence in both spectra from a science frame and a FLAT frame (see Figure \ref{fig:fringe_spectra}), it is plausible to attribute this noise to instrumental factors.
Although only the H band results are shown in this subsection, similar results were observed in the Y/J band.

\subsubsection{Not Related to an Aperture Width}

Our initial suspicion was that this noise arose from variations in pixels along the trace line of the aperture.
Because the dispersion direction is not aligned parallel with the detector, the central position of an aperture within a pixel shifts along the trace line.
To investigate this, we folded the spectrum of an aperture according to its sub-pixel position of the trace center. 
However, we did not observe recurrent patterns, which suggests that the noise is not associated with the centers of the trace lines.

\subsubsection{A Positive Correlation with the Wavelength}

We characterized this noise by measuring the period in the FLAT spectrum by using the Lomb-Scargle periodogram.
The periodogram exhibited three distinct peaks.
Notably, these peaks remained consistent across different observation frames and dates. 
However, they varied across wavelengths, with longer wavelengths (higher order numbers) exhibiting longer periodicities. 
While we have yet to identify the exact cause of this noise, its dependency on wavelength strongly suggests an instrumental origin.

\subsection{Variabilities with Observation Date and Targets}

We checked whether the periodicity of both short- and long-period noise changed or remained stable across the observation dates.
Table \ref{table:obs_speckle} lists the targets and observation dates used in this section.

\begin{table}[ht]
    \caption{Observation data used in Figures \ref{fig:fringe_spectra_June} and \ref{fig:fringe_periodogram_June}}
    \label{table:obs_speckle}
    \centering
    \begin{tabular}{ccrr}
        \hline \hline
        Target & Date & Exp. time [s] & Frames \\ \hline
        HR~7672~B & 2021-06-06 & 600 & 4 \\
            & 2021-06-07 & 600 & 5 \\
            & 2021-06-22 & $600+300$ & $3+1$ \\
            & 2021-06-24 & 1500 & 2 \\
        $\kappa$ And b & 2021-06-24 & 1800 & 2 \\ \hline
    \end{tabular}
\end{table}

\begin{figure*}[]
  \begin{center}
     \includegraphics[width=0.8\linewidth]{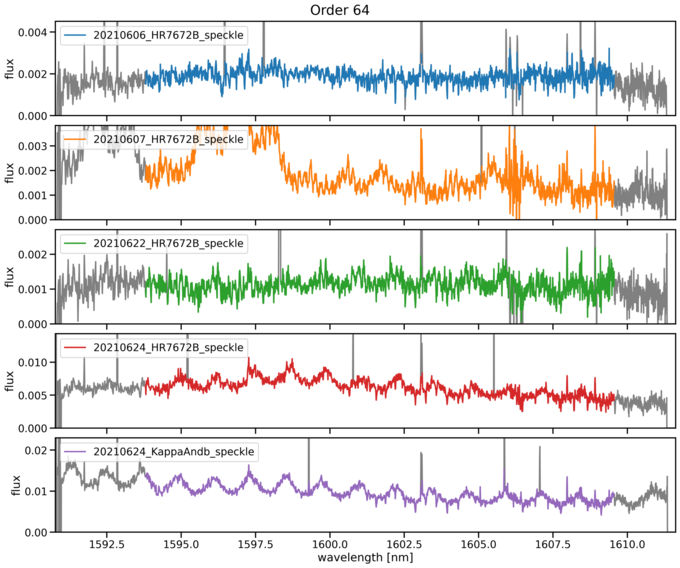}
  \end{center}
  \caption{Mean combined spectra from the speckle halo fiber in order 64 observed on different dates. From top to bottom: June 6th, 7th, 22nd, 24th for HR~7672~B, and June 24th for $\kappa$ And b. The colored points are selected by rejecting both edges of the order and clipping outliers beyond 4 sigma.
  \label{fig:fringe_spectra_June}}
\end{figure*} 
\begin{figure*}[]
  \begin{center}
     \includegraphics[width=0.8\linewidth]{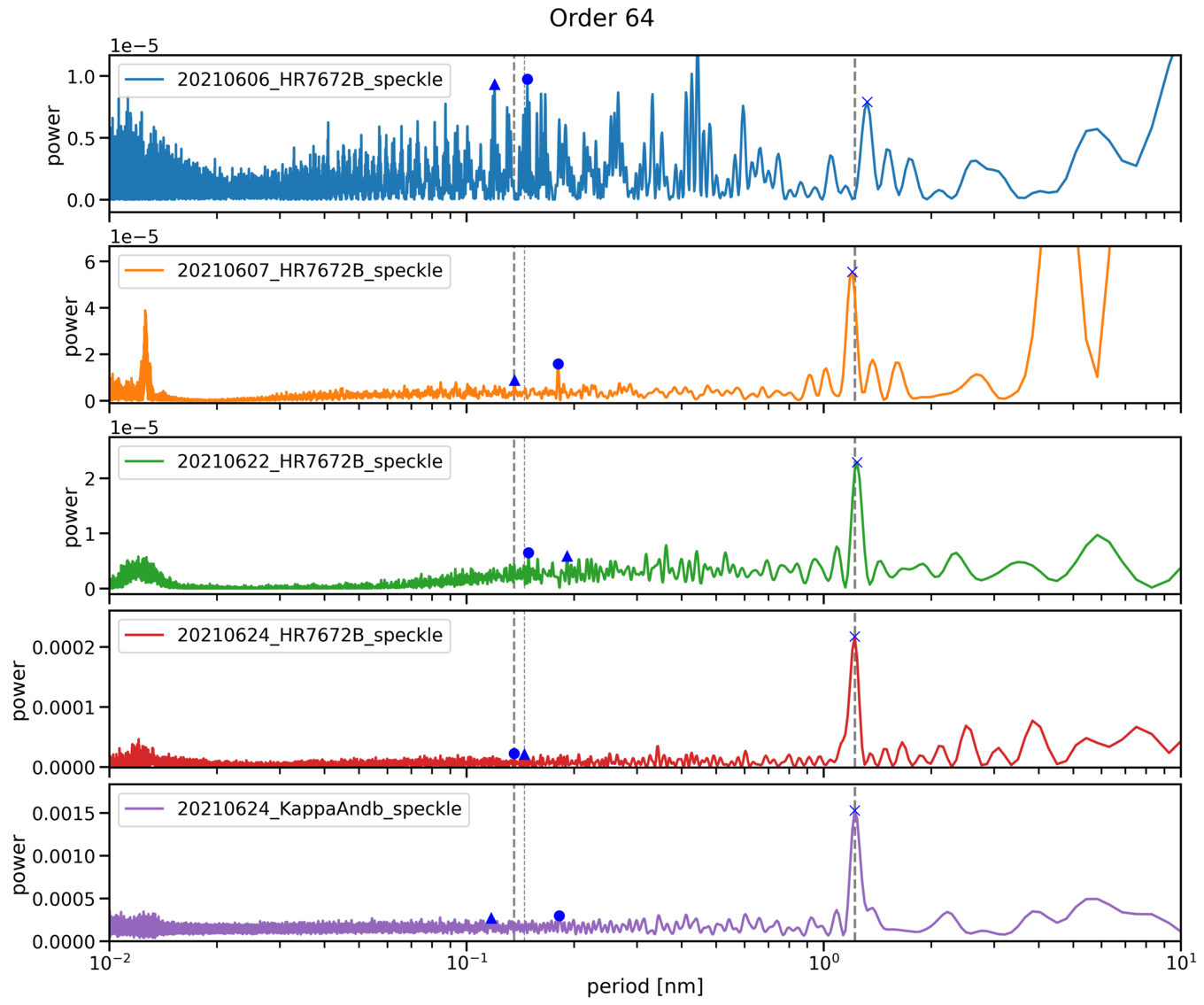}
  \end{center}
  \caption{Periodograms of spectra in order 64 of the speckle halo fiber observed on different dates, corresponding to the spectra in Figure \ref{fig:fringe_spectra_June}.
  \label{fig:fringe_periodogram_June}}
\end{figure*} 
Figures \ref{fig:fringe_spectra_June} and \ref{fig:fringe_periodogram_June} show the mean combined spectra acquired from the speckle halo fiber and their respective periodograms across various nights of observation.
Notably, the spectra of HR~7672~B and $\kappa$ And b, both observed on June 24th (as shown in the bottom two panels of each figure), exhibit similar long-term trends.
This similarity indicates that within a single night observation, variations in the noise pattern are likely small.
Importantly, the top three panels correspond to spectra with identical exposure times, suggesting that exposure time is unlikely to be responsible for the observed amplitude variations or overall trends.
However, a systematic analysis has not yet been conducted, and further investigation is needed to fully assess the potential influence of exposure time on long-period noise features.

The spectra from different nights show variations in amplitude, reaching peaks of approximately 15–22 \% (the top four panels of Figure \ref{fig:fringe_spectra_June}).
Yet, the periodogram (Figure \ref{fig:fringe_periodogram_June}) suggests that they have almost similar periodicities.
While the shorter periodicity is not prominent in order 65, the consistent short periodicity appears across the different spectra in some other orders, with a sharp peak in their periodograms.
In addition, as we mentioned in the above subsection, this short periodicity also appears in the FLAT spectra.
We have confirmed that those periods do not change regardless of the observation date (with FLAT images taken on June 7, 23, and 24, 2021).
For the longer periodicity, it features a broader peak relative to the shorter counterpart, and the period varies across orders.
Some orders show strong long-period noise, while others do not.

Our tentative conclusion based on the current dataset is that the periodicities of both short- and long-period noise remain consistent across different observation dates. 
However, the amplitudes and overall trends appear to vary. 
More data or a more detailed analysis, accounting for potential contamination from telluric lines or brown dwarf absorption features, would be needed to confirm this.

\section{Cross-correlation function analysis for molecular detection}
\label{sec:ccf_mols}

To justify the detection of \ce{FeH} in the J-band spectrum and \ce{H2O} in the H-band spectrum, as claimed in Section \ref{sec:exojax_results_feducial}, we performed a cross-correlation function (CCF) analysis, following a method similar to that described in \citet{2021A&A...656A..76Z} and \citet{10.3847/1538-4357/ac8673}. 
First, we generated spectral templates that include opacity sources of CIA, clouds, and either \ce{FeH} or \ce{H2O}, using the best-fit parameters from the mass-constrained cloudy model summarized in Table \ref{table:parameters_results}. 
These templates were not rotationally broadened.
Next, we prepared two types of residual spectra: $R = (\mathrm{observed\; data} - \mathrm{model\; without\; the\; target\; molecule})$ and $R^{\prime} = (\mathrm{observed\; data} - \mathrm{full\; model\; including\; the\; target\; molecule})$.
Then, the template spectrum was Doppler-shifted from $-1000$ to $+1000\, \mathrm{km\,s^{-1}}$ in steps of $1\, \mathrm{km\,s^{-1}}$. 
The continuum was subtracted from each spectrum using a high-pass filter.
Finally, the CCFs were normalized by the standard deviation of the CCF wings (i.e., $-1000$ to $-500$ and $+500$ to $+1000~\mathrm{km\,s^{-1}}$), allowing the result to be interpreted in terms of S/N.

We expect a prominent peak at $0\, \mathrm{km\,s^{-1}}$ (i.e., the rest frame of the model) in the CCF between the template and $R$, while no significant peak should appear in the CCF between the template and $R^{\prime}$. 
This would confirm that the inclusion of the molecule in the model explains the signal.
Wavelength regions severely affected by telluric contamination (order 57) were excluded from the analysis.

\begin{figure}[]
  \gridline{\fig{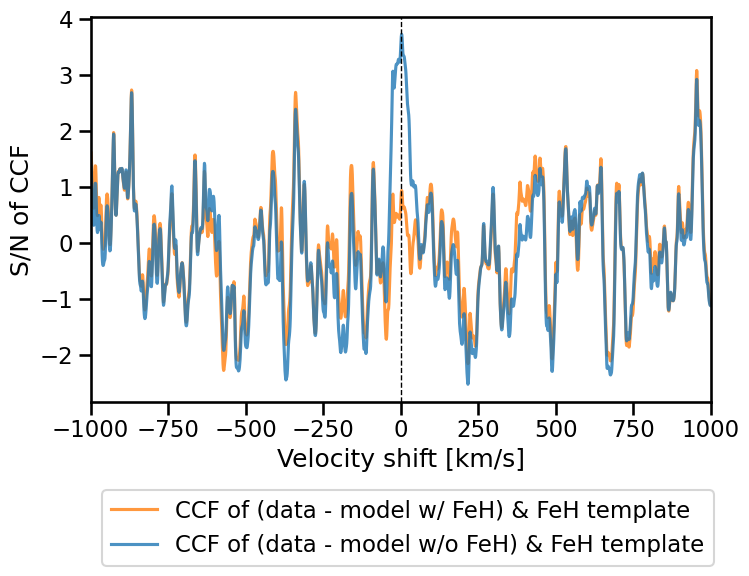}{0.9\linewidth}{(a) Cross-correlation functions for the J band}}
  \gridline{\fig{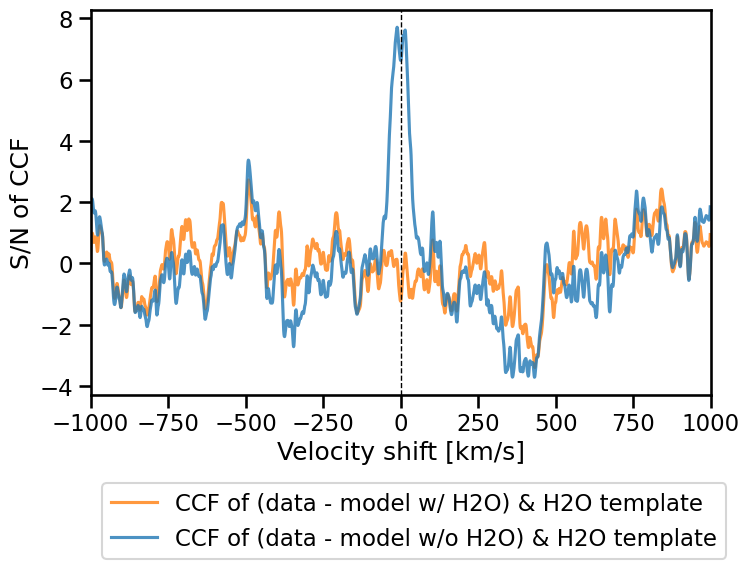}{0.9\linewidth}{(b) Cross-correlation functions for the H band }}
  \caption{Cross-correlation analysis demonstrating the detection of \ce{FeH} in the J-band spectrum in panel (a) and the detection of \ce{H2O} in the H-band spectrum in panel (b). The vertical black dashed line shows the rest frame of the brown dwarf.}
  \label{fig:ccf_mols}
\end{figure} 

Figure~\ref{fig:ccf_mols} shows the resulting CCFs.
Panels (a) and (b) demonstrate the detection of \ce{FeH} in the J-band and \ce{H2O} in the H-band, respectively, based our retrievals.
We found clear detections of molecular features as indicated by prominent CCF peaks at $0\, \mathrm{km\,s^{-1}}$ (i.e., the rest frame of the model) between the template and $R$ (blue lines), whereas no significant peak was observed in the CCFs between the template and $R^{\prime}$ (orange lines).
The S/N values at $0\, \mathrm{km\,s^{-1}}$ are 3.6 for \ce{FeH} and 6.9 for \ce{H2O}.

\section{Dependence of pressure on the contribution function}
\label{sec:cf_appendix}

To understand the behavior shown in Figure \ref{fig:cf_peaks}, we checked the pressure at which the contribution function reaches its maximum value. 
We consider the condition:
\begin{equation}
    \label{eq:dcf_dP}
    \frac{d}{dP}cf(P) = \frac{d}{dP}\left(B(\lambda,T)\frac{d e^{-\tau}}{d\log P} \right)=0
\end{equation}
for the optical depth $\tau$ associated with line opacity and CIA opacity.
For simplicity, we assume the atmosphere is isothermal, with a constant temperature at all altitudes, $T(P)=\bar{T}$ (const.).
Under this assumption, $dB(\lambda, T(P))/dP=0$, reducing Equation \eqref{eq:dcf_dP} to
\begin{equation}
    \label{eq:dcf_dP2}
    \frac{d}{dP}cf(P) = B(\lambda,T)\frac{d}{dP}\left(\frac{d e^{-\tau}}{d\log P} \right)=0.
\end{equation}
This isothermal assumption neglects the temperature dependence of both the cross-section and the CIA coefficient. 
However, their impact on the contribution function appears to be minimal.
For convenience, in this section, we denote $\sigma_i(\bar{T})$ as $\sigma_i$ and $\beta_{i,j}(\bar{T})$ as $\beta_{i,j}$.

The optical depth of the spectral lines is derived by integrating Equation \eqref{eq:opaline} from an arbitrary pressure $P$ to the pressure at the top of the atmosphere $P_\mathrm{top}$:
\begin{align}
    \label{eq:opdepth_line}
    \tau_{\mathrm{line},i} &= - \frac{x_i \sigma_i}{\mu m_u g} \int_P^{P_\mathrm{top}} dP^{\prime} \notag \\
    & = \frac{x_i \sigma_i}{\mu m_u g} (P-P_\mathrm{top}).
\end{align}
Next, using $d/d \log P = P d/dP$, we have:
\begin{align}
    \frac{d e^{-\tau_{\mathrm{line},i}}}{d \log P} 
    &= P \frac{d}{dP} \left( \exp{\left( -\frac{x_i \sigma_i}{\mu m_u g} (P-P_\mathrm{top}) \right)} \right) \notag \\
    &= -\frac{x_i \sigma_i}{\mu m_u g} P \exp{\left( -\frac{x_i \sigma_i}{\mu m_u g} (P-P_\mathrm{top}) \right)}
\end{align}
Thus, the derivative of the contribution function with respect to pressure is:
\begin{align}
    \frac{d}{dP}cf(P) &\propto \frac{d}{dP} \left( \frac{d e^{-\tau_{\mathrm{line},i}}}{d \log P} \right) \notag \\
    &= -\frac{x_i \sigma_i}{\mu m_u g} \frac{d}{dP} \left( P \exp{\left( -\frac{x_i \sigma_i}{\mu m_u g} (P-P_\mathrm{top}) \right)} \right) \notag \\
    &= -\frac{x_i \sigma_i}{\mu m_u g} \exp{\left( -\frac{x_i \sigma_i}{\mu m_u g} (P-P_\mathrm{top}) \right)} \notag \\
    & \; \times \left( 1 - \frac{x_i \sigma_i}{\mu m_u g} P\right).
\end{align}
Setting $\frac{d}{dP}cf(P)=0$ gives:
\begin{equation}
    P = \frac{\mu m_u g}{x_i \sigma_i} \propto \frac{g}{x_i}.
\end{equation}

The optical depth of CIA is derived by integrating Equation \eqref{eq:opacia}.
Here, the number density of the molecule is given by $n_i=\xi_iP/k_\mathrm{B}T$. 
Thus, we obtain:
\begin{align}
    \tau_{\mathrm{CIA},i,j} &= - \beta_{i,j} \frac{\xi_i}{k_\mathrm{B} \bar{T}} \frac{\xi_j}{k_\mathrm{B} \bar{T}} \frac{k_\mathrm{B} \bar{T}}{\mu m_u g} \int_P^{P_\mathrm{top}}P^{\prime2}\frac{1}{P^{\prime}}dP^{\prime} \notag \\
    &= - \frac{\beta_{i,j} \xi_i \xi_j}{k_\mathrm{B} \bar{T} \mu m_u g} \int_P^{P_\mathrm{top}}P^{\prime}dP^{\prime} \notag \\
    &= \frac{\beta_{i,j} \xi_i \xi_j}{k_\mathrm{B} \bar{T} \mu m_u g} \frac{1}{2} (P^2 - P_\mathrm{top}^2).
\end{align}
Then, 
\begin{align}
    \frac{d e^{-\tau_{\mathrm{CIA},i,j}}}{d \log P} 
    &= P \frac{d}{dP} \left( \exp{ \left( - \frac{\beta_{i,j} \xi_i \xi_j}{2 k_\mathrm{B} \bar{T} \mu m_u g} (P^2 - P_\mathrm{top}^2) \right)} \right) \notag \\
    &= - \frac{\beta_{i,j} \xi_i \xi_j}{k_\mathrm{B} \bar{T} \mu m_u g} P^2 \exp{ \left( - \frac{\beta_{i,j} \xi_i \xi_j}{2 k_\mathrm{B} \bar{T} \mu m_u g} (P^2 - P_\mathrm{top}^2) \right)}.
\end{align}
Therefore,
\begin{align}
    \frac{d}{dP}cf(P) &\propto \frac{d}{dP} \left( \frac{d e^{-\tau_{\mathrm{CIA},i,j}}}{d \log P} \right) \notag \\
    &= - \frac{\beta_{i,j} \xi_i \xi_j}{k_\mathrm{B} \bar{T} \mu m_u g} \notag \\
    & \; \times \frac{d}{dP}\left( P^2\exp{ \left( - \frac{\beta_{i,j} \xi_i \xi_j}{2 k_\mathrm{B} \bar{T} \mu m_u g} (P^2 - P_\mathrm{top}^2) \right)} \right) \notag \\
    &= - \frac{\beta_{i,j} \xi_i \xi_j}{k_\mathrm{B} \bar{T} \mu m_u g} P  \exp{ \left( - \frac{\beta_{i,j} \xi_i \xi_j}{2 k_\mathrm{B} \bar
    {T} \mu m_u g} (P^2 - P_\mathrm{top}^2) \right)} \notag \\
    & \; \times\left( 2 - \frac{\beta_{i,j} \xi_i \xi_j}{k_\mathrm{B} \bar{T} \mu m_u g} P^2 \right).
\end{align}
Setting $\frac{d}{dP}cf(P)=0$, and from $P \neq 0$:
\begin{equation}
    P = \left( \frac{2 k_\mathrm{B} \bar{T} \mu m_u g}{\beta_{i,j} \xi_i \xi_j} \right)^\frac{1}{2}
    \propto g^\frac{1}{2}.
\end{equation}

In summary, the dependence of the pressure at the peak of the contribution function on surface gravity is proportional to $g/x_i$ for line opacity, and to $g^{1/2}$ for CIA opacity.
If both $g$ and $x_i$ increase while keeping $g/x_i$ constant, only the latter (the contribution from CIA opacity) will shift to the lower atmosphere with higher pressure.

\bibliography{hr7672b}{}
\bibliographystyle{aasjournal}

\end{document}